\documentclass[sip,biber]{now-journal} 
\usepackage[utf8]{inputenc}
\usepackage{amssymb}
\usepackage{float}
\usepackage{amsmath,amsfonts}
\usepackage{algorithmic}
\usepackage{algorithm}
\usepackage{array}
\usepackage[caption=false,font=normalsize,labelfont=sf,textfont=sf]{subfig}
\usepackage{threeparttable}
\usepackage{textcomp}
\usepackage{stfloats}
\usepackage{url}
\usepackage{verbatim}
\usepackage{graphicx}
\usepackage{arydshln}
\usepackage{multirow} 
\usepackage{bm} 
\usepackage[normalem]{ulem}
\usepackage{chngcntr}   
\usepackage{placeins}   

\newcommand{\eg}{\emph{e.g.}}

\title{Improving Anomalous Sound Detection through Pseudo-anomalous Set Selection and Pseudo-label Utilization under Unlabeled Conditions
}

\author[1*]{Ibuki Kuroyanagi}
\author[1]{Takuya Fujimura}
\author[2]{Kazuya Takeda}
\author[3]{Tomoki Toda}

\affil[1]{The Graduate School of Informatics, Nagoya University, Aichi, Japan}
\affil[2]{The Institutes of Innovation for Future Society, Nagoya University, Aichi, Japan}
\affil[3]{The Information Technology Center, Nagoya University, Aichi, Japan}

\defbibheading{bibliography}[\refname]{}

\DeclareSourcemap{
	\maps[datatype=bibtex, overwrite=true]{
		\map{
            \step[fieldsource=booktitle,
    			match=\regexp{.*CVPR.*},
    			replace={Proc. Computer Vision and Pattern Recognition}]
            \step[fieldsource=booktitle,
    			match=\regexp{.*ICCV.*},
    			replace={Proc. International Conference on Computer Vision}]
           \step[fieldsource=booktitle,
    			match=\regexp{.*IJCNN.*},
    			replace={Proc. International Joint Conference on Neural Networks}]
		    \step[fieldsource=booktitle,
			match=\regexp{.*EUSIPCO.*},
			replace={Proc. European Signal Processing Conference}]
			\step[fieldsource=booktitle,
			match=\regexp{.*DCASE2019.*},
			replace={Proc. Detection and Classification of Acoustic Scenes and Events 2019 Workshop}]
			\step[fieldsource=booktitle,
			match=\regexp{.*DCASE2020.*},
			replace={Proc. Detection and Classification of Acoustic Scenes and Events 2020 Workshop}]
			\step[fieldsource=booktitle,
			match=\regexp{.*DCASE2021.*},
			replace={Proc. Detection and Classification of Acoustic Scenes and Events 2021 Workshop}]
			\step[fieldsource=booktitle,
			match=\regexp{.*DCASE2022.*},
			replace={Proc. Detection and Classification of Acoustic Scenes and Events 2022 Workshop}]
            \step[fieldsource=booktitle,
			match=\regexp{.*DCASE2023.*},
			replace={Proc. Detection and Classification of Acoustic Scenes and Events 2023 Workshop}]
            \step[fieldsource=booktitle,
			match=\regexp{.*DCASE2024.*},
			replace={Proc. Detection and Classification of Acoustic Scenes and Events 2024 Workshop}]
			\step[fieldsource=booktitle,
			match=\regexp{.*ICPICS.*},
			replace={Proc. International Conference on Power, Intelligent Computing and System}]
			\step[fieldsource=booktitle,
			match=\regexp{.*Computer.*Vision.*and.*Pattern.*Recognition.*},
			replace={Proc. Computer Vision and Pattern Recognition}]
			\step[fieldsource=booktitle,
			match=\regexp{.*Interspeech.*},
			replace={Proc. Interspeech}]
			\step[fieldsource=booktitle,
			match=\regexp{.*ICASSP.*},
			replace={Proc. International Conference on Acoustics, Speech and Signal Processing}]
			\step[fieldsource=booktitle,
			match=\regexp{.*International.*Conference.*on.*Learning.*Representations.*},
			replace={Proc. International Conference on Learning Representations}]
			\step[fieldsource=booktitle,
			match=\regexp{.*International.*Conference.*on.*Machine.*Learning.*},
			replace={Proc. International Conference on Machine Learning}]
			\step[fieldsource=booktitle,
			match=\regexp{.*ISIE.*},
			replace={Proc. International Symposium on Industrial Electronics}]
			\step[fieldsource=booktitle,
			match=\regexp{.*Automatic.*Speech.*Recognition.*and.*Understanding.*},
			replace={Proc. Automatic  Speech Recognition and Understanding}]
			\step[fieldsource=booktitle,
			match=\regexp{.*Spoken.*Language.*Technology.*},
			replace={Proc. Spoken Language Technology}]
			\step[fieldsource=booktitle,
			match=\regexp{.*Speech.*Synthesis.*Workshop.*},
			replace={Proc. Speech Synthesis Workshop}]
			\step[fieldsource=booktitle,
			match=\regexp{.*workshop.*on.*speech.*synthesis.*},
			replace={Proc. Speech Synthesis Workshop}]
			\step[fieldsource=booktitle, match=\regexp{.*Advances.*in.*neural.*information.*processing.*},
			replace={Proc. Neural Information Processing Systems}]
			\step[fieldsource=booktitle,
			match=\regexp{.*Advances.*in.*Neural.*Information.*Processing.*},
			replace={Proc. Neural Information Processing Systems}]
			\step[fieldsource=booktitle,
			match=\regexp{.*Workshop.*on.* Applications.* of.* Signal.*Processing.*to.*Audio.*and.*Acoustics.*},
			replace={Proc. Workshop on Applications of Signal Processing to Audio and Acoustics}]
			\step[fieldsource=series,
			match=\regexp{.+},
			replace={{}}]
			\step[fieldsource=editor,
			match=\regexp{.+},
			replace={{}}]
			\step[fieldsource=publisher,
			match=\regexp{.+},
			replace={{}}]
			\step[fieldsource=month,
			match=\regexp{.+},
			replace={{}}]
			\step[fieldsource=location,
			match=\regexp{.+},
			replace={{}}]
			\step[fieldsource=address,
			match=\regexp{.+},
			replace={{}}]
			\step[fieldsource=organization,
			match=\regexp{.+},
			replace={{}}]
		}
	}
}

\AtEveryBibitem{
  \clearfield{doi}
  \clearfield{url}
}

\issuevolumeyear{2024}
\issuevolumenumber{xx}
\issueelocation{e} 
\articletype{Original Paper} 

\articledatabox{Received xx xxxxx 2024; Revised xx xxxxx 2024\\[2pt]
ISSN 2048-7703; DOI 10.1561/116.xxxxxxxx\\
\copyright\ 2024 I. Kuroyanagi, T. Fujimura, K. Takeda and T. Toda}

\OAstatement{This is an Open Access article, distributed under the terms of the Creative Commons Attribution licence (\url{http://creativecommons.org/licenses/by-nc/4.0/}), which permits unrestricted re-use, distribution, and reproduction in any medium, for non-commercial use, provided the original work is properly cited.}


\creditline*{Corresponding author: Ibuki Kuroyanagi, kuroyanagi.ibuki@g.sp.m.is.nagoya-u.ac.jp.
This paper was partly supported by Japan's New Energy and Industrial Technology Development Organization (NEDO), project JPNP20006.
}

\begin{document}

\begin{abstract}
This paper addresses performance degradation in anomalous sound detection (ASD) when neither sufficiently similar machine data nor operational state labels are available.  
We present an integrated pipeline that combines three complementary components derived from prior work and extends them to the unlabeled‑ASD setting.
First, we adapt an anomaly‑score‑based selector to curate external audio data resembling the normal sounds of the target machine.
Second, we utilize triplet learning to assign pseudo-labels to unlabeled data, enabling finer classification of operational sounds and detection of subtle anomalies.  
Third, we employ iterative training to refine both the pseudo-anomalous set selection and pseudo-label assignment, progressively improving detection accuracy.  
Experiments on the DCASE2022--2024 Task 2 datasets demonstrate that, in unlabeled settings, our approach achieves an average AUC increase of over 6.6 points compared to conventional methods.  
In labeled settings, incorporating external data from the pseudo-anomalous set further boosts performance.  
These results highlight the practicality and robustness of our methods in scenarios with scarce machine data and labels, facilitating ASD deployment across diverse industrial settings with minimal annotation effort.
\end{abstract}

\section{Introduction}
Anomalous sound detection (ASD) systems assess whether a monitored machine is operating normally or anomalously by placing a microphone nearby and analyzing the captured audio data~\cite{ad_survey2009,Chen2018,hayashi2018anomalous,Kawaguchi2019icassp,Koizumi_DCASE2020_01,Harada_EUSIPCO2023_01}.  
Unlike anomaly detection systems relying on images or video~\cite{Ruff2020Deep,ruff2020rethinking,ding2022catching,Yao2023}, ASD systems excel in confined or dark environments inaccessible to humans, detecting anomalies through sound rather than visual inspection.  
For example, they can identify subtle issues, such as irregularities in high-speed rotating machines or minor wear in excavated areas, that cameras might miss, by capturing acoustic changes.  

Developing ASD systems is straightforward when ample anomalous data are available, but real‑world deployment faces three key challenges:  
\begin{enumerate}
    \item \textbf{Rarity of anomalous data.} Anomalous events are infrequent and diverse, prompting the use of only normal data for training to detect unknown anomalies~\cite{Koizumi_DCASE2020_01}.  
    \item \textbf{Influence of background noise.} Recorded audio comprises normal and anomalous machine sounds mixed with environmental noise, where the distinction between normal and anomalous machine sounds may be less pronounced than that between machine sounds and background noise~\cite{Wilkinghoff2024}.  
    \item \textbf{Domain shift.} Variations in machine settings and factory environments can cause mismatches between training and testing conditions, leading to misclassification of normal sounds as anomalies~\cite{Kawaguchi2021,Dohi2022_2}.  
\end{enumerate}

In the literature, unsupervised ASD usually refers to the setting in which only normal recordings are available during training and no anomalous examples are observed.
Two primary strategies address these challenges: generative and discriminative model-based methods~\cite{Kawaguchi2021}.  
Generative methods include two sub-approaches: one minimizes reconstruction errors using autoencoders~\cite{uefusa2020,Giri2020b,Purohit2020,Kapka2020,Mishra2021} or generative adversarial networks~\cite{XIA2022497,JiangAnbai2023} with normal data, while the other maximizes the likelihood of normal data using normalizing flows~\cite{NIPS2017_6c1da886,NEURIPS2020_NF,Dohi2021icassp} or Gaussian mixture models~\cite{Reynolds2009,liu2019,GuanJian2023}.  
These methods model the probability density of normal data.  
Conversely, discriminative methods classify data based on machine type and operational state (\eg, speed, location, microphone type)~\cite{stgrammfn2022,Wilkinghoff2021b,kuroyanagi2021anomalous,Chen2023wavenet,Hojjati2022,geco2023,Guan2023,choi2024noisy,Fujimura2024eu,Wilkinghoff2023eu}, deriving posterior probabilities for normal states.

Both methods solve challenge 1 (rare anomalies) by training exclusively on normal data, but they diverge on challenges 2 and 3.  
For challenge 2 (background noise), generative methods must model the entire acoustic scene and thus flag benign noise variations as anomalies, whereas discriminative methods concentrate on machine‐specific cues, yielding greater resilience to noise perturbations~\cite{Wilkinghoff2024}.  
Challenge 3 (domain shift), which manifests itself as changes in machine settings or microphone placement, poses difficulties for both methods, since shifts in the distribution of normal sounds degrade detection performance~\cite{Kawaguchi2021,Dohi2022_2}.
However, when only a handful of normal samples in the target domain are available, discriminative methods can be adapted by tweaking a few shots to recover performance, while generative methods lack this capability~\cite{Chen2022}.
Extensive benchmarks on DCASE Task 2 and related evaluations confirm these tendencies~\cite{Nishida2022eu,Wilkinghoff2023,Kuroyanagi2021dcasew}.

However, discriminative methods fail when classification tasks are overly simplistic~\cite{Koizumi_DCASE2020_01}.  
If labels are too coarse the network has the potential to capture spurious cues (e.g., noise level or band‑limited energy) instead of subtle machine abnormalities.
Consequently, performance hinges on access to similar machines and detailed state annotations during training~\cite{Wilkinghoff2024b}.  
Such labels are difficult to obtain for new equipment or inaccessible installations~\cite{Nishida2024}, leading to high false alarm rates.

Building on, yet going beyond, prior work, we make three integrative contributions: 
(i) \textbf{Pseudo‑anomalous set selection}: we extend the anomaly‑score selector of \cite{KuroyanagiNUHDL2022} by adding machine‑specific thresholds and importing the selected AudioSet clips as extra multi‑class normals (instead of binary pseudo‑anomalies). 
(ii) \textbf{Triplet‑based pseudo‑label assignment}:  we first adopt triplet learning following \cite{fujimura2024ic} to improve the fidelity of the generated pseudo-labels; building on this, we introduce a training scheme that leverages these refined pseudo-labels to achieve higher anomaly detection performance.
(iii) \textbf{Iterative learning:} we show that alternately updating the external set and the \\pseudo‑labels yields consistent AUC gains in DCASE 2022–2024 Task~2. 
To confirm the effectiveness of these methods, we conducted extensive experimental validations in unlabeled and labeled settings on DCASE 2022-2024 Task 2 datasets~\cite{Dohi2022_2,Nishida2024,Dohi2023}~\footnote{Our implementation is available at \url{https://github.com/ibkuroyagi/unlabeled-asd}.}.

This paper is structured as follows: Section~\ref{seq:2_issue} details discriminative method challenges, Section~\ref{seq:3_conventional} reviews state-of-the-art labeled approaches, Section~\ref{seq:4_Proposed} presents our method, Section~\ref{seq:5_experimental_evaluations} provides experimental results, and Section~\ref{seq6:conclusion} concludes.

\section{The Issues with Discriminative Model-based Methods}
\begin{table*}[]
\centering
\caption{Details of the datasets used in DCASE Task 2 across different years.}
\resizebox{\textwidth}{!}{
\begin{tabular}{c|cccc}
\multicolumn{1}{l|}{DCASE} & \multicolumn{1}{l}{\# of machine types} & \multicolumn{1}{l}{\# of section IDs} & Attribute labels available & \multicolumn{1}{l}{\# of training samples per machine type} \\ \hline
2022 & 7  & 6 & \checkmark & 6000 \\
2023 & 14 & 1 & \checkmark & 1000 \\
2024 & 16 & 1 & \checkmark & 1000 \\ \hline
\end{tabular}
\label{table:dataset_details}
}
\end{table*}

\label{seq:2_issue}

Discriminative model-based methods identify differences in operational sounds, such as those caused by machine manufacturers or settings, as defined by labels.  
When an anomalous sound is input, the model classifies it into a class different from its original class, thereby identifying it as an anomaly.  
Consequently, these methods are significantly influenced by the types of data in the training set and the granularity of the labels~\cite{Wilkinghoff2024b}.  

For example, in the ASD competition DCASE Challenge Task 2, datasets from 2022 to 2024 include operational sounds from 7 to 16 machine types, labeled with machine type, section ID, and attribute.
Table~\ref{table:dataset_details} shows the details of the datasets.  
The machine type indicates the kind of machine, the section ID represents domain shifts within the same machine type, and the attribute details the operational settings.  
Notably, the section ID also serves as a label for product models or manufacturers within certain machine types, making its classification equivalent to distinguishing similar sounds within a machine type.  
Prior to DCASE2022, methods that classified section IDs to detect subtle differences within machine types achieved high performance~\cite{Purohit_DCASE2019_01,Koizumi_WASPAA2019_01,Tanabe_WASPAA2021_01,Harada2021,Dohi2022_mimii}.  
From DCASE2023 onward, when section IDs were unavailable, classifying attribute labels improved performance~\cite{Nishida2024,Dohi2023}.  

However, label information suitable for ASD, such as section IDs or attributes, is not always accessible.  
For instance, when deploying an ASD system in a new factory, the monitored machines are often of the same manufacturer and product model~\cite{Dohi2023}, making it challenging to collect data equivalent to section IDs.  
Moreover, obtaining attribute labels is difficult for machines where state monitoring or setting annotation is impractical~\cite{Nishida2024}.  
Thus, existing methods struggle to achieve high performance when suitable data for ASD are unavailable.

\section{State-of-the-art Method Under Labeled Conditions}
\label{seq:3_conventional}

\subsection{Network and Input}
We adopt the architecture of Wilkinghoff~\cite{Wilkinghoff2024b}, enhanced with Fujimura’s multi‐resolution refinement~\cite{FujimuraNU2024}.  Each audio recording is first converted into three complementary representations: a 2D magnitude spectrogram obtained by applying a 256\,ms‐window short‐time Fourier transform (STFT), a second 2D magnitude spectrogram using an 8\,ms‐window STFT, and a 1D magnitude spectrum computed by a discrete Fourier transform (DFT)  over the entire signal.  Each representation is then fed into its own CNN branch, producing a 128‑dimensional embedding
\(
  \mathbf{z}^{(m)}\in\mathbb{R}^{128},\; m=1,2,3 .
\)
The three embeddings are concatenated to form 
\(
  \mathbf{z}^{\mathrm{cat}}
  =[\,\mathbf{z}^{(1)},\mathbf{z}^{(2)},\mathbf{z}^{(3)}\,]
  \in\mathbb{R}^{384}.
\)

\subsection{Sub‑cluster AdaCos and Subspace Loss}
Following~\cite{Wilkinghoff2024}, each embedding is trained with the
Sub‑cluster AdaCos (SCAC) loss, which shrinks intra‑class variance and
enlarges inter‑class margins.
Fujimura~\emph{et al.}~\cite{fujimura2024ic} introduce the subspace loss
\begin{equation}
  \mathcal{L}_{\mathrm{ss}}
  =\mathcal{L}_{\mathrm{SCAC}}(\mathbf{z}^{\mathrm{cat}},l)
   +\sum_{m=1}^{3}
     \mathcal{L}_{\mathrm{SCAC}}\!\bigl(\mathbf{z}^{(m)},l\bigr),
\end{equation}
where $l$ is the one‑hot label that combines machine type and
attribute.
The first term fixes class centres to stabilise the global embedding,
whereas the second terms encourage each sub‑space to carry discriminative cues on its own, approximating FeatEx~\cite{Wilkinghoff2024b} behaviour without explicit feature swapping.
Data augmentation uses mixup~\cite{mixupzhang2018} exactly as reported in~\cite{Wilkinghoff2024b,FujimuraNU2024}.

\subsection{Inference}
After training, source‑domain embeddings are clustered by $k_{\text{so}}$‑means
and the target‑domain ones by $k_{\text{ta}}$‑means
($k_{\text{so}}{=}16$, $k_{\text{ta}}{=}10$).

\begin{align}
  \mathcal{C}_{\mathrm{so}} &= \{\mathbf{c}_1,\dots,\mathbf{c}_{k_{\text{so}}}\},\\
  \mathcal{C}_{\mathrm{ta}} &= \{\mathbf{c}_{k_{\text{so}}+1},\dots,
                               \mathbf{c}_{k_{\text{so}}+k_{\text{ta}}}\},\\
  \mathcal{C} &= \mathcal{C}_{\mathrm{so}}\cup\mathcal{C}_{\mathrm{ta}},\quad
  J = k_{\text{so}} + k_{\text{ta}} .
\end{align}

For a test embedding $\mathbf{z}$ we compute the cosine similarity to
every representative
\begin{equation}
  s_j(\mathbf{z})
  =\frac{\langle\mathbf{z},\mathbf{c}_j\rangle}
         {\|\mathbf{z}\|\,\|\mathbf{c}_j\|},
  \qquad j=1,\dots,J .
\end{equation}
The anomaly score is then
\begin{equation}
  \mathrm{score}(\mathbf{z})
  =-\max_{1\le j\le J}\,s_j(\mathbf{z}),
\end{equation}
so that larger values indicate that $\mathbf{z}$ lies farther from
every normal cluster.%

\section{Proposed Method}
\label{seq:4_Proposed}

\begin{figure*}[!t]
\centering
\includegraphics[width=0.7\textwidth]{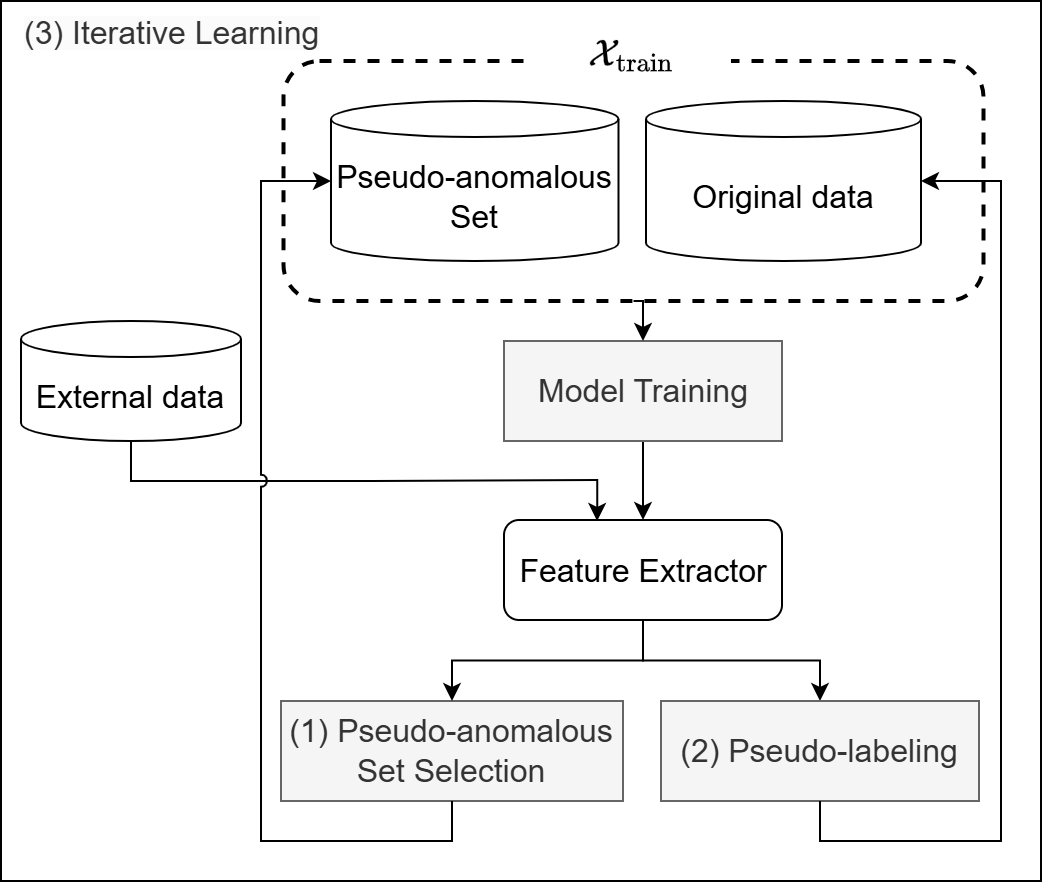}
\caption{Overview of the proposed method, illustrating the integration of three key components within an iterative learning framework: (1) Selection of the pseudo-anomalous set from external data using a feature extractor, (2) Assignment of pseudo-labels to unlabeled original data via the same feature extractor, and (3) Iterative learning, where the model is retrained over multiple cycles using updated training data \(\mathcal{X}_{\text{train}}\) derived from both the pseudo-anomalous set and original data to progressively improve performance.}
\label{fig:overview}
\end{figure*}

In unlabeled conditions where section IDs and attributes are unavailable, the performance of discriminative models significantly degrades~\cite{fujimura2024ic}. 
Building upon the subspace loss \(\mathcal{L}_{\text{ss}}\) and inference procedure described in Section~\ref{seq:3_conventional}, we propose new steps to address the lack of labels and enhance ASD performance. 
Our proposed method consists of three components:
\begin{enumerate}
    \item \textbf{Pseudo-anomalous set selection from external data:} External data similar to the normal data of the target machine type are selected based on a machine-specific threshold.
    \item \textbf{Assigning pseudo-labels to unlabeled data:} Class labels are assigned to unlabeled data using triplet learning.
    \item \textbf{Iterative learning:} The model is retrained iteratively to refine performance.
\end{enumerate}
An overview of the proposed methods is shown in Figure~\ref{fig:overview}. 
We describe each component in detail in the following subsections.

\subsection{Pseudo-anomalous Set Selection from External Data}
\label{subseq:41_ex_data}

\begin{figure*}[!t]
\centering
\includegraphics[width=0.95\textwidth]{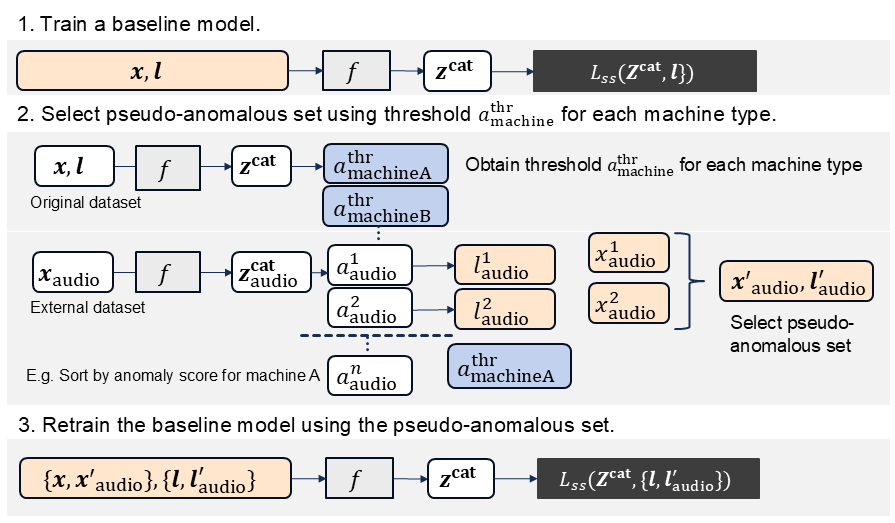}
\caption{Overview of the process for selecting the pseudo-anomalous set from external data, consisting of three steps: (1) Training a baseline model on the original dataset with the subspace loss \( \mathcal{L}_{\text{ss}}\), (2) Processing external data through the baseline model to compute anomaly scores, which are then sorted and filtered using machine-specific thresholds \( a^\text{thr}_{\text{machineA}}, a^\text{thr}_{\text{machineB}}, \ldots \) to select the pseudo-anomalous set, and (3) Retraining the baseline model on the combined dataset to refine predictions with the subspace loss \( \mathcal{L}_{\text{ss}}\).}
\label{fig:use_external_data}
\end{figure*}

The performance of discriminative models decreases when class labels representing similar sounds within the same machine type (\eg, section IDs) are unavailable. 
The proposed method addresses this by selecting external data that resemble the target machine type's normal sounds and assigning appropriate class labels. 
An overview of this process is shown in Figure~\ref{fig:use_external_data}. 
The pseudo-anomalous set selection consists of three steps: (i) training a baseline model, (ii) selecting the pseudo-anomalous set from external data using a machine-specific threshold, and (iii) retraining the baseline model with the pseudo-anomalous set. 
We adopt the baseline method proposed by Fujimura~\cite{FujimuraNU2024} to train the model for selecting external data.

\subsubsection{Selecting the Pseudo-anomalous Set from External Data}

To mitigate the performance drop, the proposed method selects external data that are misclassified as normal by the baseline model. 
The trained baseline model computes anomaly scores for all training data, and for each machine type, the highest anomaly score is used as a threshold, denoted as \(a^{\text{thr}}_{\text{machine}}\). 
External data with anomaly scores below \(a^{\text{thr}}_{\text{machine}}\) are considered similar to the normal data of the corresponding machine type. 
To prevent over-reliance on external data, the number of external samples added per machine type is capped at \(N_{\text{max}}\). 
Specifically, if \(N_{\text{out}}\) is the number of selected external samples, the number of samples added is:
\begin{equation}
N_{\text{ex}} = \min(N_{\text{out}}, N_{\text{max}}),
\end{equation}
where the \(N_{\text{ex}}\) external samples with the smallest anomaly scores are added to the training data. 
This expanded dataset is referred to as \(\mathcal{X}_{\text{train}}\). 
The classification labels for the external data follow the format \(\texttt{machine}\_{\texttt{attribute}}\), where:
\begin{itemize}
    \item \(\texttt{machine}\) is assigned based on the machine type from the original dataset, using the class with the highest similarity to the external dataset.
    \item \(\texttt{attribute}\) is assigned from the external dataset's class label (if available).
\end{itemize}
If multiple external samples belong to the same class but are associated with different machine types, they are treated as separate classes to account for the potentially coarser class definitions in external data.

Conventional methods using external data~\cite{KuroyanagiNUHDL2022,primus2020anomalous} randomly select external data, define them as pseudo-anomalous, and train the model using binary classification. 
However, these methods may select irrelevant sounds (\eg, instruments or speech), limiting their effectiveness. 
Our method improves performance by filtering external data that are beneficial for ASD and treating them as part of a multi-class classification problem, enabling the detection of subtle differences among normal data.

\subsubsection{Retraining the Model with the Pseudo-anomalous Set}

The feature extractor is retrained on the augmented dataset \(\mathcal{X}_{\text{train}}\) using the same procedure as the baseline. 
After training, representative vectors are calculated only from the original dataset, excluding external data, to prevent misclassifying anomalous sounds similar to external data as normal. 
The methods for calculating anomaly scores and representative vectors remain the same as in the baseline~\cite{FujimuraNU2024}.

\subsection{Assigning Pseudo-labels to Unlabeled Data}
\label{subseq:42_plabel}

\begin{figure*}[!t]
\centering
\includegraphics[width=0.8\textwidth]{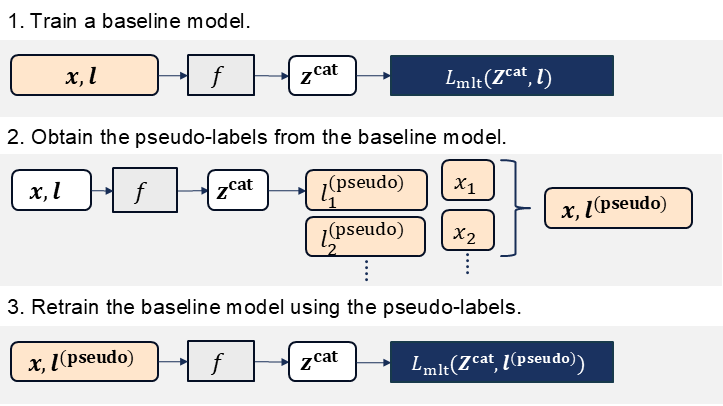}
\caption{Overview of the proposed method for assigning pseudo-labels to unlabeled data, consisting of three steps: (1) Training a baseline model on the original dataset by passing it through the function \( f \) to produce category outputs \( z^{\text{cat}} \), optimized using the loss function \( \mathcal{L}_{\text{mlt}}\), (2) Using the trained baseline model to predict and assign pseudo-labels to unlabeled data, generating \( x, l^{(\text{pseudo})} \), and (3) Retraining the model on the pseudo-labeled data \( x, l^{(\text{pseudo})} \) to iteratively improve predictions via the loss function \( \mathcal{L}_{\text{mlt}}\).}
\label{fig:use_pseudo_label}
\end{figure*}

We propose a method to improve performance when the machine type is known, but internal state or configuration labels (\eg, \(\texttt{attribute}\)) are unavailable. 
An overview of this method is shown in Figure~\ref{fig:use_pseudo_label}. 
The process consists of three steps: (i) training a baseline model, (ii) obtaining pseudo-labels from the baseline model, and (iii) retraining the baseline model with the pseudo-labels.

\subsubsection{Training a Baseline Model}
\label{subsubseq:triplet}
In unlabeled conditions where attribute labels are unavailable, baseline training typically relies solely on machine type labels. 
However, this approach often leads to a simplistic classification task, causing the model to focus on irrelevant features such as specific frequencies or background noise~\cite{Wilkinghoff2024}, which can impair its ability to detect subtle anomalies.
To address this limitation and prepare the model for effective pseudo-labeling, we enhance the baseline training by incorporating triplet learning, building on insights from prior work~\cite{fujimura2024ic}. 
Specifically, \cite{fujimura2024ic} demonstrated that triplet learning can effectively disentangle operational sound variations from environmental noise, creating a feature space better suited for anomaly detection. 
We selected triplet learning instead of contrastive learning because it better aligns with our goal of capturing subtle changes within the same sample, such as variations in machine operational sounds, rather than just distinguishing between different samples.
Contrastive learning focuses on separating distinct samples within a mini-batch, which is less effective for identifying fine-grained differences critical to anomaly detection. 
In contrast, triplet learning encourages the model to emphasize these subtle intra-sample changes while ignoring irrelevant noise variations. 

For triplet learning, we define three samples: the anchor \(\bm{x}_i^a\), the positive \(\bm{x}_i^p\), and the negative \(\bm{x}_i^n\), as follows:
\begin{itemize}
    \item \textbf{Anchor}: The anchor sample \(\bm{x}_i^a\) is a normal sound sample from the \(i\)-th machine type.
    \item \textbf{Positive}: The positive sample \(\bm{x}_i^p\) is created by adding scaled sound from a different machine type \(j \neq i\) as background noise to the anchor, defined as:
    \begin{equation}
        \bm{x}_i^p = \bm{x}_i^a + 10^{-\frac{\alpha}{20}} \cdot \frac{\|\bm{x}_i^a\|}{\|\bm{x}_j\|} \bm{x}_j,
    \end{equation}
    where \(\|\cdot\|\) denotes the Euclidean norm, and \(\alpha\) is a hyperparameter representing the signal-to-noise ratio (SNR) in decibels to adjust the intensity of the background noise relative to the anchor signal.
    \item \textbf{Negative}: The negative sample \(\bm{x}_i^n\) is generated by pitch-shifting the anchor sample:
    \begin{equation}
        \bm{x}_i^n = \text{PitchShift}(\bm{x}_i^a, \beta),
    \end{equation}
    where \(\beta\) simulates operational variations, with the implementation based on torchaudio~\cite{torchaudio2022,torchaudio2023}.
\end{itemize}
This triplet configuration encourages the model to distinguish operational changes from background noise variations. Let \(\bm{z}_i^a\), \(\bm{z}_i^p\), and \(\bm{z}_i^n\) represent the embedding vectors of \(\bm{x}_i^a\), \(\bm{x}_i^p\), and \(\bm{x}_i^n\), respectively.

A similarity function with a temperature parameter \(\tau\) is employed:
\begin{equation}
\label{eq:sim_tau}
s_{\tau}(\bm{z}, \bm{z}') = \frac{\langle \bm{z}, \bm{z}' \rangle}{\tau \, \|\bm{z}\| \|\bm{z}'\|},
\end{equation}
where \(\langle \cdot, \cdot \rangle\) denotes the dot product. The triplet loss \(\mathcal{L}_{\text{trp}}\) is defined as:
\begin{equation}
\label{eq:triplet_loss_final}
\mathcal{L}_{\text{trp}} (\bm{z}_i^a, \bm{z}_i^p, \bm{z}_i^n) = \max \left\{0, \gamma + 1 - s_{\tau}(\bm{z}_i^a, \bm{z}_i^p) + s_{\tau}(\bm{z}_i^a, \bm{z}_i^n) \right\},
\end{equation}
where \(\gamma\) is a margin parameter.
This loss encourages the model to prioritize operational sound differences over noise.

The feature extractor is trained using both the triplet loss \(\mathcal{L}_{\text{trp}}\) and the subspace loss \(\mathcal{L}_{\text{ss}}\), where \(\bm{l}_i\) is the one-hot machine type label. Mixup~\cite{mixupzhang2018} is applied to \(\mathcal{L}_{\text{ss}}\) with a 50\,\% probability, but not to \(\mathcal{L}_{\text{trp}}\), to preserve the intended triplet relationships. The combined training loss is:
\begin{equation}
\label{eq:L_mlt}
\mathcal{L}_{\text{mlt}} = \mathcal{L}_{\text{trp}}(\bm{z}_i^a, \bm{z}_i^p, \bm{z}_i^n) + \mathcal{L}_{\text{ss}}(\bm{z}_i, \bm{l}_i).
\end{equation}
This enhanced baseline training establishes a robust feature space, enabling effective pseudo-label assignment in the subsequent steps and supporting the iterative learning framework.

\subsubsection{Generating Pseudo-labels}
\label{subsubseq:plabel}

After training, we generate pseudo-labels by applying \(k\)-means clustering to the learned embeddings for each domain and machine type, similar to previous studies~\cite{fujimura2024ic, JiangCUP2024, LvAITHU2024}. 
Let \(k_{\text{so}}\) and \(k_{\text{ta}}\) be the numbers of clusters for the source and target domains, respectively. 
Each sample \(\bm{x}_i\) is assigned to the cluster with the nearest centroid in embedding space:
\begin{equation}
\label{eq:pseudo_label_assign}
\ell_i^{(\mathrm{pseudo})} = \underset{1 \le j \le k_{d_i}}{\operatorname{arg\,min}} \left\| \bm{z}_i - \bm{c}_j^{(d_i)} \right\|, \quad \text{where } d_i = \begin{cases} \text{so} & \text{(source domain)}, \\ \text{ta} & \text{(target domain)}. \end{cases}
\end{equation}

\subsubsection{Retraining the Baseline Model with Pseudo-labels}

We retrain the model from scratch using \(\mathcal{L}_{\text{mlt}}\), replacing \(\bm{l}_i\) with the pseudo-label \(\ell_i^{(\text{pseudo})}\). 
Since pseudo-labels are not ground truth, minimizing intra-class variance with \(\mathcal{L}_{\text{ss}}\) alone may force samples with different true labels into the same cluster, leading the model to focus on environmental noise or incorrectly group operational sounds. 
By incorporating the triplet loss, the model learns to ignore noise differences and focus on operational sound changes, mitigating these issues and improving ASD performance. 
For inference, the procedure remains the same as the baseline.

\subsection{Iterative Selection of Pseudo-anomalous Set and Pseudo-labels}
\label{subseq:43_iterate}

We combine the pseudo-anomalous set selection (Section~\ref{subseq:41_ex_data}) and pseudo-label assignment (Section~\ref{subseq:42_plabel}) into an iterative training scheme:
\begin{itemize}
    \item[\(\bullet\)] \textbf{Stage 1:} Train the model using \(\mathcal{L}_{\text{mlt}}\) on the original labeled dataset (no external data or pseudo-labels).
    \item[\(\bullet\)] \textbf{Stage \(M\) (\(M \ge 2\)):} 
    \begin{enumerate}
        \item Use the model from stage \((M{-}1)\) to select the pseudo-anomalous set and add up to \(N_{\text{max}}\) samples per machine type to the training set.
        \item Use the same model to assign pseudo-labels via \(k\)-means clustering.
        \item Retrain the model using the augmented dataset and new pseudo-labels with \(\mathcal{L}_{\text{mlt}}\).
    \end{enumerate}
\end{itemize}
An overview of this method is shown in Figure~\ref{fig:overview}. 
By iterating, we refine the model at each stage, potentially obtaining more accurate pseudo-anomalous sets and meaningful pseudo-labels for the next stage. 
During inference, we reuse the procedure summarised in Section~\ref{seq:3_conventional}:
for each test embedding we take the negative of its maximum cosine
similarity to the representative vectors obtained from source and target domain training data.
This iterative approach progressively enhances performance by leveraging both external data and pseudo-labeled internal variations.

\section{Experimental Evaluations}
\label{seq:5_experimental_evaluations}
\subsection{Datasets}
\label{subseq:51_dataset}
We evaluated our proposed methods using the DCASE Task 2 datasets from 2022 to 2024~\cite{Dohi2022_2,Nishida2024,Dohi2023}. 
These datasets comprise machine sound recordings, including factory background noise. 
Each recording is a single-channel audio file, lasting 6--18 seconds, with a 16\,kHz sampling rate. 
The DCASE2022 dataset includes seven machine types: fan, gearbox, bearing, slide rail (slider), valve, ToyCar, and ToyTrain~\cite{Harada2021,Dohi2022}. 
The DCASE2023 dataset features 14 machine types: its development set matches DCASE2022, while the evaluation set includes ToyDrone, ToyTank, ToyNscale, bandsaw, grinder, and shaker~\cite{Harada2023}. 
The DCASE2024 dataset contains 16 machine types: its development set aligns with DCASE2022, and the evaluation set includes 3D-Printer, AirCompressor, BrushlessMotor, HairDryer, HoveringDrone, RoboticArm, Scanner, ToothBrush, and ToyCircuit~\cite{Albertini2024,Niizumi2024}. 
Table~\ref{table:dataset_details} summarizes the label conditions for each dataset. 
Note that some attribute labels in DCASE2024, unavailable during the competition, were released post-competition and used in this analysis.

Each dataset provides 1,000 training samples per machine type, all normal data. 
These consist of 990 source-domain samples and 10 target-domain samples affected by domain shift. 
Domain shift occurs due to changes in machine sound characteristics, such as those caused by maintenance actions or variations in the acoustic environment (\eg, background noise or shifts in operating conditions).
Training samples include attribute labels indicating the machine's operating state or environment. 
An ideal ASD system should detect anomalies reliably despite domain shifts without adaptation~\cite{Wang2023}. 
Per the DCASE Task 2 setup, training data indicate source or target domain origin. 
Evaluation data include 100 normal and 100 anomalous samples per machine type, split evenly between domains. 
During inference, the domain of test data is unknown. 
In DCASE2022, section IDs denote domain shift types, used alongside attribute labels in labeled settings. 
From DCASE2023 onward, the absence of section IDs simplified the classification task, making it harder to extract embeddings sensitive to anomalous changes.

Performance is assessed using the area under the ROC curve (AUC), following DCASE Task 2. 
The AUC is vital as machine condition monitoring thresholds aim to minimize false alarms~\cite{auc_Donna,auc_Ebbers}. 
This threshold-independent metric offers an objective comparison of ASD systems.
\subsection{System Descriptions}
\label{subseq:52_system_description}

In this study, we compare our proposed method in an unlabeled configuration with Wilkinghoff's method, known for state-of-the-art performance across multiple datasets in conventional ASD systems~\cite{Wilkinghoff2024b}, and Fujimura's baseline method~\cite{FujimuraNU2024}. 
Since ASD hyperparameters cannot be tuned with anomalous data, we use identical hyperparameter values for all machine types. 
This approach, widely adopted in ASD~\cite{Dohi2022_2,Nishida2024,Dohi2023}, ensures robust performance on unseen machine types. 
We outlined the settings for each method below.

\textbf{Wilkinghoff's Method (Wilkinghoff~\cite{Wilkinghoff2024b}).} 
This method, per~\cite{Wilkinghoff2024b}, uses a magnitude spectrogram and the full magnitude spectrum as input features. 
Feature extraction employs a window size of 64\,ms with a 50\,\% hop size. 
Two convolutional branches produce 128-dimensional embedding vectors, concatenated into a 256-dimensional feature for ASD. 
The SCAC loss~\cite{Wilkinghoff2024} is applied to 16 untrainable sub-clusters (randomly initialized) using these 256-dimensional features. 
Training involved a batch size of 100, 50 epochs, the AdamW optimizer~\cite{adamw2019} with a learning rate of 0.001, and mixup with uniformly sampled ratios. 
Post-training, embeddings are clustered via $k$-means: 16 clusters for the source domain and 10 for the target domain, where each target-domain sample serves as its own representative vector. 
The optimal cluster number in DCASE2023 was 16 for the source domain, though its impact was minor~\cite{Wilkinghoff2024b}. 
Anomaly scores are derived from cosine similarity between test samples and representative vectors.

\textbf{Baseline (\texttt{Ba}~\cite{FujimuraNU2024}).} 
This baseline adopts most settings from Wilkinghoff~\cite{Wilkinghoff2024b}, with key differences: 
(i) it uses two magnitude spectrograms (8\,ms and 256\,ms windows) plus the full magnitude spectrum, all with 50\,\% hop sizes; 
(ii) three convolutional branches each yield a 128-dimensional embedding, concatenated to form a 384-dimensional feature; 
(iii) SCAC loss~\cite{Wilkinghoff2024} is applied twice, first to 16 untrainable sub-clusters for the 384-dimensional feature, then to 16 trainable sub-clusters per 128-dimensional branch (all randomly initialized). 
Other training and inference parameters align with Wilkinghoff's method.

\textbf{Baseline with External Data (\texttt{Ba}+\texttt{Ex}).} 
Following Section~\ref{subseq:41_ex_data}, we incorporate approximately 1.8 million labeled audio samples from Audioset~\cite{audioset} as external data. 
Each Audioset sample carries a ``mid'' label (\eg, \texttt{/m/05r5c} for Piano, \texttt{/m/05zppz} for Male speech), used as attribute labels. 
Representative vectors are computed from the training set using $k$-means with 16 clusters for the source domain. 
Up to $N_{\text{max}}=1000$ external samples per machine type, matching the training data size, are selected. 
The baseline model is retrained with these labeled samples, retaining the original hyperparameters.

\textbf{Baseline with \(\mathcal{L}_{\text{trp}}\) (\texttt{Ba}+\(\mathcal{L}_{\text{trp}}\)).} 
Following Section~\ref{subsubseq:triplet}, this method enhances the baseline with triplet learning.
Positive samples are generated by adding scaled sound from a different machine type as background noise, with the hyperparameter \(\alpha\) (SNR, \([-5, 20] \, \text{dB}\)) controlling noise intensity.
The range of \([-5, 20] \, \text{dB}\) was chosen to balance the intensity of the noise and the original sound source, ensuring that the noise-to-signal ratio remains within a range where both components are comparable.
Negative samples are created by pitch shifting the anchor, with \(\beta\) ranging from \(\pm 6\) to \(\pm 12\) semitones; this implementation is based on torchaudio~\cite{torchaudio2022,torchaudio2023}.
The range of \(\beta\) was selected to simulate realistic operational variations without excessively distorting the original sound, covering a semitone shift to an octave shift.
The triplet loss uses \(\tau = 0.2\) and \(\gamma = 0.5\), following the configuration in~\cite{Kuroyanagiicassp2022}.

\textbf{Baseline with Pseudo-Labels (\texttt{Ba}+\texttt{Ps}).} 
Per Section~\ref{subsubseq:plabel}, this method retrains the baseline using pseudo-labels. 
After initial training, embeddings are clustered per domain: 16 clusters ($k_{\text{so}}=16$) for the source domain and 4 ($k_{\text{ta}}=4$) for the target domain per machine type. 
Assuming 16 source-domain attributes and 4 target-domain attributes, this yields 20 pseudo-classes per machine type. 
Samples are assigned pseudo-labels based on the nearest of 20 centroids. 
From stage~2, the model will address a classification task with 20 times the classes of stage~1's machine types.

\textbf{Iterative Learning Method (\texttt{Ba}+\(\mathcal{L}_{\text{trp}}\)+\texttt{Ps}+\texttt{Ex}, Stage \(M\)).} 
Per Section~\ref{subseq:43_iterate}, this iterative method boosts performance over up to \(M=5\) iterations. 
From stage \(M=3,4,5\), pseudo-labels and external data are derived using the model from stage \(M-1\).

\subsection{Performance Evaluation When Labels are Unavailable }
\label{subseq:57_label_unable}
\begin{table*}[]
\centering
\caption{Average AUC (\%) for each training configuration on the DCASE 2022--2024 Task~2 datasets.  
The upper block (``w/ label'') is a label‑based reference, while the lower block (``w/o label'') shows the proposed unlabeled configurations. 
Column ``stage’’ indicates the iteration number: stage 1 is the initial model, stage 2 adds external data and\,/\,or pseudo‑labels derived from stage 1, and stages 3–5 repeat the same update procedure recursively.  
Columns ``dev’’ and ``eval’’ correspond to development and evaluation splits.  
Values are mean $\pm$ variance over five random seeds.  
\texttt{Ba}\,$=$\,baseline~\cite{FujimuraNU2024}, \texttt{Ex}\,$=$\,selected external data, \texttt{Ps}\,$=$\,pseudo‑labels, and $\mathcal{L}_{\text{trp}}=$\,triplet loss.}

\label{table:1_unlabel}
\resizebox{\textwidth}{!}{
\begin{tabular}{c|l|c|cccccc}
\multirow{2}{*}{Use label} &
  \multicolumn{1}{c|}{\multirow{2}{*}{Method}} &
  \multirow{2}{*}{stage} &
  \multicolumn{2}{c}{2022} &
  \multicolumn{2}{c}{2023} &
  \multicolumn{2}{c}{2024} \\ \cline{4-9} 
 &
  \multicolumn{1}{c|}{} &
   &
  dev &
  eval &
  dev &
  eval &
  dev &
  eval \\ \hline
\multirow{2}{*}{w/ label} &
  Wilkinghoff~\cite{Wilkinghoff2024b} &
  1 &
  82.5$\pm$0.8 &
  73.1$\pm$0.9 &
  67.2$\pm$0.8 &
  74.2$\pm$0.3 &
  72.6$\pm$0.7 &
  61.5$\pm$0.6 \\
 &
  \texttt{Ba}~\cite{FujimuraNU2024} &
  1 &
  81.9$\pm$0.9 &
  73.0$\pm$0.3 &
  70.5$\pm$0.5 &
  77.5$\pm$0.4 &
  72.8$\pm$0.4 &
  63.2$\pm$0.8 \\ \cline{1-9}
\multirow{9}{*}{w/o label} &
  Wilkinghoff~\cite{Wilkinghoff2024b} &
  1 &
  67.2$\pm$5.8 &
  64.1$\pm$0.8 &
  62.5$\pm$0.8 &
  64.4$\pm$3.2 &
  59.7$\pm$1.1 &
  55.6$\pm$0.6 \\
 &
  \texttt{Ba}~\cite{FujimuraNU2024} &
  1 &
  71.3$\pm$0.9 &
  64.8$\pm$0.7 &
  64.2$\pm$1.2 &
  67.8$\pm$1.4 &
  59.5$\pm$0.7 &
  53.8$\pm$0.6 \\
 &
  \texttt{Ba}+\(\mathcal{L}_{\text{trp}}\) &
  1 &
  71.8$\pm$1.6 &
  65.2$\pm$1.3 &
  64.1$\pm$1.2 &
  68.8$\pm$0.6 &
  59.7$\pm$1.3 &
  54.1$\pm$1.5 \\
 &
  \texttt{Ba}+\(\mathcal{L}_{\text{trp}}\)+\texttt{Ex} &
  2 &
  74.0$\pm$0.5 &
  65.3$\pm$1.3 &
  65.0$\pm$1.1 &
  69.2$\pm$0.3 &
  61.2$\pm$1.3 &
  54.9$\pm$0.9 \\
 &
  \texttt{Ba}+\(\mathcal{L}_{\text{trp}}\)+\texttt{Ps} &
  2 &
  76.2$\pm$0.4 &
  68.4$\pm$0.8 &
  64.2$\pm$1.3 &
  72.4$\pm$0.7 &
  65.5$\pm$1.5 &
  56.4$\pm$1.1 \\
 &
  \texttt{Ba}+\(\mathcal{L}_{\text{trp}}\)+\texttt{Ps}+\texttt{Ex} &
  2 &
  75.8$\pm$0.9 &
  69.1$\pm$0.7 &
  64.4$\pm$0.5 &
  72.7$\pm$1.0 &
  66.4$\pm$2.2 &
  56.5$\pm$0.6 \\
 &
  \texttt{Ba}+\(\mathcal{L}_{\text{trp}}\)+\texttt{Ps}+\texttt{Ex} &
  3 &
  76.8$\pm$1.7 &
  70.1$\pm$1.4 &
  \textbf{65.2$\pm$0.5} &
  72.6$\pm$1.3 &
  68.3$\pm$1.0 &
  \textbf{57.0$\pm$0.1} \\
 &
  \texttt{Ba}+\(\mathcal{L}_{\text{trp}}\)+\texttt{Ps}+\texttt{Ex} &
  4 &
  76.5$\pm$2.0 &
  70.1$\pm$0.6 &
  64.3$\pm$1.9 &
  \textbf{73.1$\pm$0.6} &
  \textbf{70.7$\pm$1.5} &
  56.1$\pm$0.0 \\
 &
  \texttt{Ba}+\(\mathcal{L}_{\text{trp}}\)+\texttt{Ps}+\texttt{Ex} &
  5 &
  \textbf{78.1$\pm$1.0} &
  \textbf{70.3$\pm$1.1} &
  \textbf{65.2$\pm$1.4} &
  72.6$\pm$0.3 &
  70.4$\pm$1.5 &
  56.8$\pm$0.1 \\ \hline
\end{tabular}
}
\end{table*}
Table~\ref{table:1_unlabel} presents the experimental results evaluating the performance of the proposed methods under unlabeled conditions. 
The per-domain performance is reported in Appendix Table~\ref{table:appendix_unlabel}. 
To assess the effect of \(\mathcal{L}_{\text{trp}}\), we compare \texttt{Ba} and \texttt{Ba}+\(\mathcal{L}_{\text{trp}}\), focusing on the ``all'' column. 
Incorporating \(\mathcal{L}_{\text{trp}}\) improved performance on all datasets except DCASE2024 development, where the difference was minimal. 
This suggests that \(\mathcal{L}_{\text{trp}}\) generally enhances performance, likely by suppressing noise and emphasizing variations in operating sounds within samples, creating a more discriminative feature space.

To verify the effect of retraining with external data, we compared \texttt{Ba}+\(\mathcal{L}_{\text{trp}}\) and \texttt{Ba}+\(\mathcal{L}_{\text{trp}}\)+\texttt{Ex}, focusing on the ``all'' column. 
Using external data improved performance across all datasets, demonstrating its effectiveness. 
Specifically, adding external data similar to normal data appears to enhance the detection of anomalies resembling those external samples.

To examine the effect of retraining with pseudo-labels, we compared \texttt{Ba}+\(\mathcal{L}_{\text{trp}}\) and \texttt{Ba}+\(\mathcal{L}_{\text{trp}}\)+\texttt{Ps}, focusing on the ``all'' column. 
Retraining with pseudo-labels improved performance on all datasets, indicating its effectiveness. 
Assigning samples with similar operating sounds to the same class enables the detection of finer differences, such as those across machine settings, leading to performance gains.

To determine whether combining external data and pseudo-labels yields additional benefits, we compared \texttt{Ba}+\(\mathcal{L}_{\text{trp}}\)+\texttt{Ex}, \texttt{Ba}+\(\mathcal{L}_{\text{trp}}\)+\texttt{Ps}, and \texttt{Ba}+\(\mathcal{L}_{\text{trp}}\)+\texttt{Ps}+\texttt{Ex}, focusing on the ``all'' column. 
In most datasets, the combined method outperformed each individual method, except in DCASE2022 development (where \texttt{Ba}+\(\mathcal{L}_{\text{trp}}\)+\texttt{Ps} performed best) and DCASE2024 development (where \texttt{Ba}+\(\mathcal{L}_{\text{trp}}\)+\texttt{Ex} performed best). 
In these exceptions, target-domain performance was notably high, suggesting that particularly suitable external data or pseudo-labels were obtained. 
Since combining both methods never degraded performance relative to each single method across all datasets, we consider their joint use effective.

We next evaluated the performance improvements from iterative learning by comparing \texttt{Ba}+\(\mathcal{L}_{\text{trp}}\)+\texttt{Ps}+\texttt{Ex} at stages 2, 3, 4, and 5, focusing on the ``all'' column. 
The average scores across all datasets increased from 67.5 at stage 2 to 68.3, 68.5, and 68.9 at stages 3, 4, and 5, respectively, indicating steady improvement. 
Performance improved in both source and target domains, demonstrating that the proposed method effectively boosts performance under unlabeled conditions. 
For individual datasets, stages 3, 4, and 5 outperformed stage 2, but differences among stages 3, 4, and 5 were not pronounced. 
Given that anomalous data are unavailable for model validation, iterating up to stage 3 appears sufficient for developing ASD systems for unknown machines.

Finally, we compared our unlabeled results with the upper-bound performance using ground-truth labels. 
The average ``all''-column scores across all datasets for w/ label Wilkinghoff~\cite{Wilkinghoff2024b}, w/ label \texttt{Ba}~\cite{FujimuraNU2024}, w/o label Wilkinghoff~\cite{Wilkinghoff2024b}, and w/o label \texttt{Ba}~\cite{FujimuraNU2024} are 71.9, 73.2, 62.3, and 63.6, respectively. 
In \texttt{Ba}~\cite{FujimuraNU2024}, the gap between labeled and unlabeled settings was 9.4 points, but our approach at stage 5 reduced this gap to 4.3 points, improving performance by 5.1 points. 
Moreover, our stage 5 result is only 3.0 points below the w/ label Wilkinghoff~\cite{Wilkinghoff2024b} score. 
This significant reduction in the performance gap shows that our method greatly improves upon existing unlabeled approaches and closely approaches label-based performance.

These results indicate that, in an unlabeled setting, using \(\mathcal{L}_{\text{trp}}\), external data, and pseudo-labeling with at least three iterations is effective.

\begin{table*}[]
\centering
\caption{Average AUC (\%) of each supervised configuration (attribute labels available) on the DCASE 2022–2024 Task~2 datasets.  
Column ``stage’’ indicates the training iteration: stage 1 is the initial model; stage 2 retrains the baseline with external data (Ex) and\,/\,or pseudo‑labels (Ps) derived from stage 1; stage 3 repeats the same update.  
Columns ``dev’’ and ``eval’’ correspond to the development and evaluation splits, respectively (mean $\pm$ variance over five random seeds).  
\texttt{Ba}\,$=$\,baseline~\cite{FujimuraNU2024}, $\mathcal{L}_{\text{trp}}\,=$\,triplet loss.}

\label{table:1_label}
\resizebox{\textwidth}{!}{
\begin{tabular}{l|c|cccccc}
\multicolumn{1}{c|}{\multirow{2}{*}{Method}} &
  \multirow{2}{*}{stage} &
  \multicolumn{2}{c}{2022} &
  \multicolumn{2}{c}{2023} &
  \multicolumn{2}{c}{2024} \\ \cline{3-8} 
\multicolumn{1}{c|}{}                    &   & dev                   & eval                  & dev          & eval                  & dev          & eval                  \\ \hline
Wilkinghoff~\cite{Wilkinghoff2024b}      & 1 & \textbf{82.5$\pm$0.8} & 74.2$\pm$0.3          & 73.1$\pm$0.9 & 72.6$\pm$0.7          & 67.2$\pm$0.8 & 61.5$\pm$0.6          \\
\texttt{Ba}~\cite{FujimuraNU2024}        & 1 & 81.9$\pm$0.9          & \textbf{77.5$\pm$0.4} & 73.0$\pm$0.3 & 72.8$\pm$0.4          & 70.5$\pm$0.5 & 63.2$\pm$0.8          \\
\texttt{Ba}+\(\mathcal{L}_{\text{trp}}\) & 1 & 80.7$\pm$0.8          & 68.8$\pm$0.3          & 72.6$\pm$2.3 & 69.0$\pm$0.9          & 68.8$\pm$0.9 & 61.9$\pm$1.6          \\
\texttt{Ba}+\texttt{Ex} &
  2 &
  81.7$\pm$0.3 &
  76.9$\pm$0.5 &
  \textbf{73.8$\pm$1.0} &
  \textbf{75.0$\pm$0.9} &
  \textbf{71.2$\pm$0.9} &
  \textbf{64.4$\pm$0.6} \\
\texttt{Ba}+\texttt{Ps}                  & 2 & 79.8$\pm$0.4          & 76.0$\pm$0.4          & 73.6$\pm$0.6 & 71.7$\pm$1.6          & 70.2$\pm$0.6 & 61.0$\pm$1.1          \\
\texttt{Ba}+\texttt{Ps}+\(\mathcal{L}_{\text{trp}}\) &
  2 &
  80.0$\pm$0.5 &
  75.8$\pm$0.5 &
  72.5$\pm$0.6 &
  69.0$\pm$1.6 &
  69.1$\pm$0.9 &
  60.7$\pm$0.8 \\
\texttt{Ba}+\texttt{Ex}                  & 3 & 82.2$\pm$0.8          & 76.8$\pm$0.5          & 73.4$\pm$0.5 & \textbf{75.0$\pm$1.5} & 70.2$\pm$1.1 & \textbf{64.4$\pm$1.2} \\ \hline
\end{tabular}
}
\end{table*}
\subsection{Performance Evaluation When Labels are Available}
\label{subseq:56_label_able}

We evaluated the proposed method in a labeled setting using the results presented in Table~\ref{table:1_label}.  
The per-domain performance is reported in Appendix Table~\ref{table:appendix_label}. 
The pseudo-label approach assumes that original labels lack sufficient granularity for ASD.  
It augments these labels with pseudo-labels to enable finer-grained classification.  
Specifically, when applying pseudo-labels, the label format shifts from \texttt{machine\_attribute} to \texttt{machine\_attribute\_pseudo-label}.

We compared \texttt{Ba}~\cite{FujimuraNU2024} and \texttt{Ba}+\(\mathcal{L}_{\text{trp}}\) to evaluate triplet learning's impact.
When ground-truth labels are available, incorporating \(\mathcal{L}_{\text{trp}}\) degrades performance across all datasets.  
Since \(\mathcal{L}_{\text{trp}}\) drives the model to detect subtle intra-sample variations, it may overemphasize minor sound differences when original labels already adequately represent machine settings.  
This oversensitivity likely explains the performance decline.

Next, we examined the effect of external data by comparing \texttt{Ba}~\cite{FujimuraNU2024} with \texttt{Ba}+\texttt{Ex} in the ``all'' column.  
Given that \(\mathcal{L}_{\text{trp}}\) does not enhance performance in this setting, we use \texttt{Ba}~\cite{FujimuraNU2024} as the stage 1 model.  
Adding external data improves performance across all DCASE2023 and DCASE2024 datasets.  
However, no significant improvement occurs in DCASE2022.  
When using external data, samples similar to the normal data are selected from external sources.  
In settings without predefined machine type distinctions (\eg, unlabeled conditions) or similar machine sounds (\eg, lacking section IDs), external data increase classification complexity, potentially boosting performance.  
Conversely, when section IDs are present, the training data already contain ample similar operational sounds, limiting the impact of external data on task complexity and thus yielding minimal gains.

We then evaluated pseudo-labeling by comparing \texttt{Ba}~\cite{FujimuraNU2024} with \texttt{Ba}+\texttt{Ps} in the ``all'' column.  
Pseudo-labels degrade performance in all datasets except the DCASE2023 development set.  
In a labeled setting, pseudo-labeling effectively subdivides existing labels into finer categories.  
If original labels accurately reflect machine settings, further subdivision fragments the data unnecessarily.  
These additional categories may capture irrelevant variations (\eg, background noise), rendering pseudo-labeling redundant when the original data are well-segmented.

Finally, we compared stage 2 and stage 3 of \texttt{Ba}+\texttt{Ex} to assess iterative learning benefits.  
Since neither \(\mathcal{L}_{\text{trp}}\) nor pseudo-labeling proves advantageous in this setting, we iterate using \texttt{Ba}+\texttt{Ex}.  
Results indicate negligible improvement from stage 2 to stage 3.  
In stage 2, the model learns to distinguish Audioset samples from normal data based on stage 1 selections.  
Without new insights from \(\mathcal{L}_{\text{trp}}\) or pseudo-labeling, re-extracting external data from Audioset offers no additional perspective, leaving performance largely unchanged.

These findings suggest that in the labeled setting where the original labels are appropriately annotated, applying the external data approach once is the most effective strategy.
\subsection{Effectiveness of Proposed External Data Selection and Impact of External Data Volume}
\label{subseq:58_ablation1}
\begin{table*}[]
\centering
\caption{Performance evaluation comparing external data selection via anomaly scores (proposed method) and random selection, with varying maximum numbers of external data ($N_{\max}$). 
\textit{Large $N_{\mathrm{out}}$ machines} refers to machine types where $N_{\mathrm{out}} \geq N_{\max}$, 
while \textit{small $N_{\mathrm{out}}$ machines} refers to those where $N_{\mathrm{out}} < N_{\max}$. 
Each value represents the average AUC [\%] across all machine types in the dataset, 
with variance computed from five runs using different random seeds.}
\label{table:3_n_out}
\resizebox{\textwidth}{!}{
\centering
\begin{tabular}{l|r|ccc|ccc}
\multirow{2}{*}{} &
  \multicolumn{1}{l|}{\multirow{2}{*}{$N_\mathrm{max}$}} &
  \multicolumn{3}{c|}{\textit{large $N_{\mathrm{out}}$ machines}} &
  \multicolumn{3}{c}{\textit{small $N_{\mathrm{out}}$ machines}} \\ \cline{3-8} 
 &
  \multicolumn{1}{l|}{} &
  2022 &
  2023 &
  2024 &
  2022 &
  2023 &
  2024 \\ \hline
\texttt{Ba}~\cite{FujimuraNU2024} &
  \multicolumn{1}{c|}{--} &
  71.8$\pm$1.2 &
  58.4$\pm$0.6 &
  59.9$\pm$2.3 &
  67.3$\pm$0.7 &
  65.6$\pm$0.5 &
  59.2$\pm$0.6 \\
\texttt{Ba}+\texttt{Ex} &
  500 &
  74.1$\pm$1.2 &
  61.2$\pm$1.9 &
  65.1$\pm$2.7 &
  68.3$\pm$0.6 &
  \textbf{66.6$\pm$0.8} &
  59.3$\pm$0.8 \\
\texttt{Ba}+\texttt{Ex} &
  1000 &
  \textbf{74.6$\pm$0.8} &
  \textbf{62.3$\pm$2.2} &
  \textbf{66.1$\pm$5.7} &
  68.7$\pm$1.1 &
  66.3$\pm$1.5 &
  \textbf{60.0$\pm$0.8} \\
\texttt{Ba}+\texttt{Ex} &
  2000 &
  74.1$\pm$0.9 &
  60.6$\pm$2.5 &
  63.8$\pm$5.4 &
  \textbf{69.0$\pm$0.6} &
  66.3$\pm$0.6 &
  \textbf{60.0$\pm$1.0} \\
\texttt{Ba}+\texttt{Ex} (random) &
  500 &
  73.3$\pm$0.9 &
  60.3$\pm$2.4 &
  63.6$\pm$2.4 &
  67.8$\pm$0.9 &
  66.1$\pm$1.2 &
  59.8$\pm$0.8 \\
\texttt{Ba}+\texttt{Ex} (random) &
  1000 &
  73.5$\pm$1.1 &
  60.6$\pm$1.9 &
  59.2$\pm$2.9 &
  67.4$\pm$0.3 &
  65.8$\pm$1.0 &
  59.3$\pm$0.6 \\
\texttt{Ba}+\texttt{Ex} (random) &
  2000 &
  74.1$\pm$0.7 &
  61.5$\pm$1.5 &
  61.1$\pm$1.1 &
  67.6$\pm$0.5 &
  65.1$\pm$0.5 &
  58.9$\pm$1.5 \\ \hline
\end{tabular}
}
\end{table*}
We investigated the effect of pseudo-anomalous data from external data in the unlabeled settings.
Table~\ref{table:3_n_out} is analyzed in three key aspects:
\begin{enumerate}
    \item Performance differences between selecting external data via anomaly scores (our method) and random selection,
    \item Performance variations based on the maximum number of external samples, $N_\mathrm{max}$, used in training,
    \item Performance differences between machine types with $N_\mathrm{out} \geq N_\mathrm{max}$ (\textit{large $N_\mathrm{out}$ machines}) and those with $N_\mathrm{out} < N_\mathrm{max}$ (\textit{small $N_\mathrm{out}$ machines}).
\end{enumerate}

In DCASE2022, \textit{large $N_\mathrm{out}$ machines} are fan and valve, while \textit{small $N_\mathrm{out}$ machines} are bearing, gearbox, slider, ToyCar, and ToyTrain. In DCASE2023, \textit{large $N_\mathrm{out}$ machines} are bandsaw and grinder, and \textit{small $N_\mathrm{out}$ machines} include bearing, fan, gearbox, shaker, slider, ToyCar, ToyDrone, ToyNscale, ToyTank, ToyTrain, Vacuum, and valve. For DCASE2024, \textit{large $N_\mathrm{out}$ machines} are BrushlessMotor, and \textit{small $N_\mathrm{out}$ machines} are 3DPrinter, AirCompressor, bearing, fan, gearbox, HairDryer, HoveringDrone, RoboticArm, Scanner, slider, ToothBrush, ToyCar, ToyCircuit, ToyTrain, and valve.

To compare our method with random selection, we evaluated \texttt{Ba}~\cite{FujimuraNU2024} against \texttt{Ba}+\texttt{Ex}. \texttt{Ba}+\texttt{Ex} consistently improves performance across all $N_\mathrm{max}$ values, whereas \texttt{Ba}+\texttt{Ex} (random) shows performance drops in some datasets compared to \texttt{Ba}~\cite{FujimuraNU2024}. Notably, the highest performance across all datasets is achieved with our proposed method rather than random selection.
This suggests that targeting external samples likely to be misclassified as normal enhances ASD performance more effectively than random selection.
Furthermore, this improvement occurs regardless of the classification into \textit{large $N_\mathrm{out}$ machines} or \textit{small $N_\mathrm{out}$ machines}, indicating that adding external data resembling not only the target machine but also co-trained machine types contributes to the performance gain.

Next, we examined the relationship between $N_\mathrm{max}$ and performance in \texttt{Ba}+\texttt{Ex}. For \textit{large $N_\mathrm{out}$ machines}, performance decreases when $N_\mathrm{max}$ is changed from 1000 to either 500 or 2000. This indicates that while incorporating more external data prone to misclassification as normal can improve performance, excessive external data beyond the original training data leads to performance degradation. This highlights the need to balance the volume of external data with the original training data. For \textit{small $N_\mathrm{out}$ machines}, performance shows no consistent trend with varying $N_\mathrm{max}$. Since $N_\mathrm{out} < N_\mathrm{max}$, increasing $N_\mathrm{max}$ introduces external data resembling other machine types rather than the target class, which likely explains the lack of clear performance shifts.

These results demonstrate that our method, which prioritizes external data prone to misclassification as normal, outperforms random selection and effectively boosts performance, particularly for machine types with abundant relevant external data (\textit{large $N_\mathrm{out}$ machines}). Furthermore, for these machine types, setting $N_\mathrm{max}$ to a value comparable to the size of the training data (\eg, $N_\mathrm{max} = 1000$) is critical to prevent performance degradation due to excessive external data.

\subsection{Performance Analysis of Triplet Loss and Pseudo-Labels in Stage 2}
\label{subseq:5_ablation2}
\begin{table*}[]
\centering
\caption{Evaluation of the impact of the triplet loss \(\mathcal{L}_{\text{trp}}\) on pseudo-label quality and model performance in a two-stage training framework.
Columns indicate the model used in stage 1 to generate pseudo-labels \(\text{Ba}\), \(\text{Ba} + \mathcal{L}_{\text{trp}}\) and the dataset (DCASE2022, DCASE2023, DCASE2024).
Rows show the model configuration in stage 2: \(\text{Ba} + \text{Ps}\) and \(\text{Ba} + \mathcal{L}_{\text{trp}} + \text{Ps}\) . Each entry reports the average AUC (\%) across all machine types in the respective dataset, with mean and variance computed from five runs.}
\label{table:4_loss}
\resizebox{\textwidth}{!}{
\begin{tabular}{l|ccc|ccc}
Loss functions in stage 1 & \multicolumn{3}{c|}{\texttt{Ba}}           & \multicolumn{3}{c}{\texttt{Ba}+\(\mathcal{L}_{\text{trp}}\)} \\ \hline
DCASE                     & 2022         & 2023         & 2024         & 2022               & 2023               & 2024               \\ \hline
\texttt{Ba}+\texttt{Ps}   & 70.6$\pm$0.2 & 63.2$\pm$1.1 & 57.4$\pm$1.2 & 71.3$\pm$0.5       & 63.5$\pm$1.3       & 58.9$\pm$0.6       \\
\texttt{Ba}+\(\mathcal{L}_{\text{trp}}\)+\texttt{Ps} &
  \textbf{71.8$\pm$0.4} &
  \textbf{66.8$\pm$1.4} &
  \textbf{59.1$\pm$0.5} &
  \textbf{74.3$\pm$0.5} &
  \textbf{67.0$\pm$1.0} &
  \textbf{60.3$\pm$1.2} \\ \hline
\end{tabular}
}
\end{table*}

Table~\ref{table:4_loss} evaluates the performance of models trained in stage 2 using pseudo-labels generated in stage 1 by \texttt{Ba} or \texttt{Ba} + \(\mathcal{L}_{\text{trp}}\).
The results demonstrate that pseudo-labels generated by \texttt{Ba} + \(\mathcal{L}_{\text{trp}}\) in stage 1 consistently yield higher performance than those generated by \texttt{Ba} alone, regardless of the loss function used in stage 2.
This suggests that incorporating \(\mathcal{L}_{\text{trp}}\) in stage 1 enhances the quality of pseudo-labels.

Moreover, when comparing stage 2 configurations, \texttt{Ba} + \(\mathcal{L}_{\text{trp}}\) + \text{Ps} outperforms \texttt{Ba} + \text{Ps} across all datasets and pseudo-label data. 
This indicates that adding \(\mathcal{L}_{\text{trp}}\) in stage 2 reduces the negative impact of incorrectly assigned pseudo-labels.
The triplet loss contributes to both improved pseudo-label quality and reduced misclassification errors by training the model to focus on features relevant to ASD, such as changes in machine sounds, while ignoring noise and trivial machine-specific characteristics.

These findings highlight the effectiveness of incorporating triplet loss in all stages of an unsupervised learning framework with pseudo-labels, leading to enhanced model performance.
\section{Conclusion}
\label{seq6:conclusion}
This paper introduced three methods to enhance ASD performance when detailed operational state labels and similar machine data are limited.  
First, we proposed a pseudo-anomalous set selection method to address scenarios with scarce comparable machine types.  
By scanning a vast external dataset, we automatically extracted audio samples resembling the target machine's normal sounds, increasing classification complexity without compromising key characteristics.  
Second, we developed a pseudo-label assignment strategy for unlabeled data, enabling the detection of subtle operational differences critical for ASD.  
Clustering learned embedding vectors subdivided unlabeled data into pseudo-classes, refining the model's focus on fine-grained anomalies.  
Third, we implemented iterative learning to refine these techniques, recalculating anomaly scores and improving pseudo-label precision across cycles, progressively boosting detection accuracy.

Experiments were conducted in both unlabeled and labeled settings.  
In the unlabeled setting, our approach significantly outperformed conventional methods reliant on coarse machine type labels.  
In the labeled setting, incorporating selected external data further improved detection accuracy.
Incorporating external data selected based on similarity to the target's normal sounds consistently enhances detection accuracy, with our targeted selection method proving superior to random selection.  
For machine types with abundant relevant external data (\textit{large $N_{\mathrm{out}}$ machines}), setting the maximum external data volume ($N_{\mathrm{max}}$) to a value comparable to the training data size, such as $N_{\mathrm{max}} = 1000$, optimizes performance, while excessive data leads to degradation.  
Additionally, employing triplet loss in both training stages improves pseudo-label quality in stage~1 and mitigated the impact of label errors in stage~2, demonstrating its effectiveness in unsupervised learning frameworks.

Collectively, these findings confirmed that our methods robustly improved ASD performance under the limited label availability and for novel machine types, providing practical and effective solutions for industrial applications.

\appendix                           
\counterwithin{table}{section}
\section{Appendix} 
\begin{table*}[]
\centering
\caption{Average AUC (\%) of each method under unlabeled (attribute‑free) conditions on the DCASE 2022–2024 Task~2 datasets.  “source’’ and “target’’ denote the two domains; values are mean $\pm$ variance over five random seeds.}

\label{table:appendix_unlabel}
\resizebox{\textwidth}{!}{
\begin{tabular}{r|c|l|c|cccc}
\multicolumn{1}{c|}{\multirow{2}{*}{DCASE}} &
  \multirow{2}{*}{Use label} &
  \multicolumn{1}{c|}{\multirow{2}{*}{Method}} &
  \multirow{2}{*}{stage} &
  \multicolumn{2}{c}{development} &
  \multicolumn{2}{c}{evaluation} \\ \cline{5-8} 
\multicolumn{1}{c|}{} &
   &
  \multicolumn{1}{c|}{} &
   &
  source &
  target &
  source &
  target \\ \hline
\multirow{11}{*}{2022} &
  \multirow{2}{*}{w/ label} &
  Wilkinghoff~\cite{Wilkinghoff2024b} &
  1 &
  \textbf{86.0$\pm$0.9} &
  78.2$\pm$0.7 &
  77.7$\pm$0.8 &
  71.6$\pm$1.0 \\
 &
   &
  \texttt{Ba}~\cite{FujimuraNU2024} &
  1 &
  84.9$\pm$0.6 &
  78.6$\pm$1.7 &
  \textbf{80.2$\pm$0.6} &
  \textbf{74.2$\pm$1.0} \\ \cline{2-8} 
 &
  \multirow{9}{*}{w/o label} &
  Wilkinghoff~\cite{Wilkinghoff2024b} &
  1 &
  69.6$\pm$6.1 &
  64.2$\pm$5.3 &
  66.9$\pm$5.5 &
  63.0$\pm$2.2 \\
 &
   &
  \texttt{Ba}~\cite{FujimuraNU2024} &
  1 &
  71.5$\pm$1.2 &
  71.1$\pm$1.2 &
  70.4$\pm$1.6 &
  66.2$\pm$1.4 \\
 &
   &
  \texttt{Ba}+\(\mathcal{L}_{\text{trp}}\) &
  1 &
  72.1$\pm$1.8 &
  71.7$\pm$1.4 &
  70.5$\pm$1.6 &
  67.1$\pm$0.9 \\
 &
   &
  \texttt{Ba}+\(\mathcal{L}_{\text{trp}}\)+\texttt{Ex} &
  2 &
  73.6$\pm$1.0 &
  74.9$\pm$0.8 &
  71.8$\pm$0.5 &
  67.4$\pm$1.1 \\
 &
   &
  \texttt{Ba}+\(\mathcal{L}_{\text{trp}}\)+\texttt{Ps} &
  2 &
  79.5$\pm$1.0 &
  \textbf{75.1$\pm$0.9} &
  75.5$\pm$0.6 &
  \textbf{69.6$\pm$1.0} \\
 &
   &
  \texttt{Ba}+\(\mathcal{L}_{\text{trp}}\)+\texttt{Ps}+\texttt{Ex} &
  2 &
  79.0$\pm$1.1 &
  73.3$\pm$1.3 &
  76.4$\pm$1.4 &
  68.8$\pm$1.2 \\
 &
   &
  \texttt{Ba}+\(\mathcal{L}_{\text{trp}}\)+\texttt{Ps}+\texttt{Ex} &
  3 &
  80.8$\pm$2.0 &
  72.7$\pm$2.1 &
  \textbf{76.7$\pm$0.8} &
  68.5$\pm$2.3 \\
 &
   &
  \texttt{Ba}+\(\mathcal{L}_{\text{trp}}\)+\texttt{Ps}+\texttt{Ex} &
  4 &
  80.2$\pm$2.2 &
  72.8$\pm$1.8 &
  76.5$\pm$0.4 &
  69.2$\pm$0.8 \\
 &
   &
  \texttt{Ba}+\(\mathcal{L}_{\text{trp}}\)+\texttt{Ps}+\texttt{Ex} &
  5 &
  \textbf{82.9$\pm$0.6} &
  74.5$\pm$1.8 &
  76.3$\pm$0.7 &
  68.9$\pm$0.6 \\ \hline
\multirow{11}{*}{2023} &
  \multirow{2}{*}{w/ label} &
  Wilkinghoff~\cite{Wilkinghoff2024b} &
  1 &
  71.2$\pm$1.6 &
  75.0$\pm$1.5 &
  75.5$\pm$0.8 &
  68.7$\pm$2.2 \\
 &
   &
  \texttt{Ba}~\cite{FujimuraNU2024} &
  1 &
  72.0$\pm$1.4 &
  74.7$\pm$1.5 &
  78.0$\pm$1.5 &
  68.3$\pm$2.1 \\ \cline{2-8} 
 &
  \multirow{9}{*}{w/o label} &
  Wilkinghoff~\cite{Wilkinghoff2024b} &
  1 &
  64.9$\pm$1.8 &
  63.6$\pm$0.7 &
  62.8$\pm$0.8 &
  56.7$\pm$1.7 \\
 &
   &
  \texttt{Ba}~\cite{FujimuraNU2024} &
  1 &
  65.7$\pm$1.5 &
  63.5$\pm$1.2 &
  60.6$\pm$0.9 &
  57.3$\pm$1.7 \\
 &
   &
  \texttt{Ba}+\(\mathcal{L}_{\text{trp}}\) &
  1 &
  64.6$\pm$2.5 &
  64.8$\pm$1.4 &
  60.8$\pm$1.0 &
  57.8$\pm$2.5 \\
 &
   &
  \texttt{Ba}+\(\mathcal{L}_{\text{trp}}\)+\texttt{Ex} &
  2 &
  65.7$\pm$1.7 &
  64.2$\pm$1.1 &
  62.0$\pm$1.5 &
  58.9$\pm$2.2 \\
 &
   &
  \texttt{Ba}+\(\mathcal{L}_{\text{trp}}\)+\texttt{Ps} &
  2 &
  69.5$\pm$2.0 &
  67.5$\pm$1.2 &
  63.1$\pm$1.9 &
  67.9$\pm$2.3 \\
 &
   &
  \texttt{Ba}+\(\mathcal{L}_{\text{trp}}\)+\texttt{Ps}+\texttt{Ex} &
  2 &
  70.0$\pm$1.3 &
  67.5$\pm$2.5 &
  65.8$\pm$1.5 &
  66.2$\pm$4.2 \\
 &
   &
  \texttt{Ba}+\(\mathcal{L}_{\text{trp}}\)+\texttt{Ps}+\texttt{Ex} &
  3 &
  71.1$\pm$1.5 &
  \textbf{69.3$\pm$2.3} &
  65.9$\pm$0.5 &
  70.3$\pm$1.7 \\
 &
   &
  \texttt{Ba}+\(\mathcal{L}_{\text{trp}}\)+\texttt{Ps}+\texttt{Ex} &
  4 &
  71.8$\pm$1.2 &
  68.3$\pm$1.8 &
  68.6$\pm$1.6 &
  \textbf{73.3$\pm$1.0} \\
 &
   &
  \texttt{Ba}+\(\mathcal{L}_{\text{trp}}\)+\texttt{Ps}+\texttt{Ex} &
  5 &
  \textbf{72.0$\pm$1.7} &
  68.3$\pm$1.1 &
  \textbf{68.9$\pm$0.8} &
  72.7$\pm$2.0 \\ \hline
\multirow{11}{*}{2024} &
  \multirow{2}{*}{w/ label} &
  Wilkinghoff~\cite{Wilkinghoff2024b} &
  1 &
  68.9$\pm$1.3 &
  63.8$\pm$1.8 &
  63.2$\pm$1.4 &
  62.7$\pm$1.7 \\
 &
   &
  \texttt{Ba}~\cite{FujimuraNU2024} &
  1 &
  74.6$\pm$1.1 &
  64.5$\pm$1.8 &
  64.0$\pm$1.4 &
  65.3$\pm$2.4 \\ \cline{2-8} 
 &
  \multirow{9}{*}{w/o label} &
  Wilkinghoff~\cite{Wilkinghoff2024b} &
  1 &
  65.9$\pm$1.3 &
  58.8$\pm$1.4 &
  54.2$\pm$1.4 &
  57.4$\pm$0.9 \\
 &
   &
  \texttt{Ba}~\cite{FujimuraNU2024} &
  1 &
  66.1$\pm$3.2 &
  59.5$\pm$1.8 &
  52.5$\pm$1.5 &
  54.6$\pm$0.8 \\
 &
   &
  \texttt{Ba}+\(\mathcal{L}_{\text{trp}}\) &
  1 &
  65.4$\pm$1.9 &
  59.0$\pm$1.0 &
  52.6$\pm$2.1 &
  55.6$\pm$1.6 \\
 &
   &
  \texttt{Ba}+\(\mathcal{L}_{\text{trp}}\)+\texttt{Ex} &
  2 &
  68.2$\pm$2.4 &
  \textbf{61.4$\pm$0.9} &
  54.7$\pm$1.1 &
  55.2$\pm$0.8 \\
 &
   &
  \texttt{Ba}+\(\mathcal{L}_{\text{trp}}\)+\texttt{Ps} &
  2 &
  68.8$\pm$1.8 &
  59.6$\pm$0.9 &
  57.0$\pm$1.8 &
  56.5$\pm$1.3 \\
 &
   &
  \texttt{Ba}+\(\mathcal{L}_{\text{trp}}\)+\texttt{Ps}+\texttt{Ex} &
  2 &
  68.2$\pm$1.4 &
  59.8$\pm$1.4 &
  58.6$\pm$1.1 &
  55.6$\pm$0.9 \\
 &
   &
  \texttt{Ba}+\(\mathcal{L}_{\text{trp}}\)+\texttt{Ps}+\texttt{Ex} &
  3 &
  \textbf{71.6$\pm$0.0} &
  58.2$\pm$1.1 &
  58.5$\pm$1.0 &
  56.2$\pm$0.1 \\
 &
   &
  \texttt{Ba}+\(\mathcal{L}_{\text{trp}}\)+\texttt{Ps}+\texttt{Ex} &
  4 &
  69.9$\pm$1.0 &
  59.2$\pm$2.7 &
  55.7$\pm$0.9 &
  \textbf{57.5$\pm$0.7} \\
 &
   &
  \texttt{Ba}+\(\mathcal{L}_{\text{trp}}\)+\texttt{Ps}+\texttt{Ex} &
  5 &
  69.9$\pm$2.2 &
  59.9$\pm$0.4 &
  \textbf{59.1$\pm$1.4} &
  56.9$\pm$0.7 \\ \hline
\end{tabular}
}
\end{table*}

\begin{table*}[]
\centering
\caption{Average AUC (\%) of each method under labeled (attribute‑available) conditions on the DCASE 2022–2024 Task~2 datasets.  Notation is identical to Table~\ref{table:appendix_unlabel}.}
\label{table:appendix_label}
\resizebox{\textwidth}{!}{
\begin{tabular}{c|l|c|cccc}
\multirow{2}{*}{DCASE} & \multicolumn{1}{c|}{\multirow{2}{*}{Method}} & \multirow{2}{*}{stage} & \multicolumn{2}{c}{development}      & \multicolumn{2}{c}{evaluation} \\ \cline{4-7} 
 &
  \multicolumn{1}{c|}{} &
   &
  source &
  target &
  source &
  target \\ \hline
\multirow{7}{*}{2022}  & Wilkinghoff~\cite{Wilkinghoff2024b}          & 1                      & \textbf{86.0$\pm$0.9} & 78.2$\pm$0.7 & 77.7$\pm$0.8   & 71.6$\pm$1.0  \\
 &
  \texttt{Ba}~\cite{FujimuraNU2024} &
  1 &
  84.9$\pm$0.6 &
  78.6$\pm$1.7 &
  \textbf{80.2$\pm$0.6} &
  \textbf{74.2$\pm$1.0} \\
 &
  \texttt{Ba}+\(\mathcal{L}_{\text{trp}}\) &
  1 &
  83.7$\pm$0.6 &
  76.4$\pm$1.1 &
  71.2$\pm$0.5 &
  66.2$\pm$0.5 \\
 &
  \texttt{Ba}+\texttt{Ex} &
  2 &
  84.7$\pm$0.4 &
  \textbf{79.4$\pm$0.7} &
  79.9$\pm$0.8 &
  73.3$\pm$0.8 \\
 &
  \texttt{Ba}+\texttt{Ps} &
  2 &
  82.9$\pm$0.3 &
  76.4$\pm$1.4 &
  79.8$\pm$0.1 &
  72.4$\pm$1.0 \\
 &
  \texttt{Ba}+\texttt{Ps}+\(\mathcal{L}_{\text{trp}}\) &
  2 &
  82.7$\pm$0.7 &
  77.6$\pm$0.8 &
  79.1$\pm$0.5 &
  72.4$\pm$0.5 \\
 &
  \texttt{Ba}+\texttt{Ex} &
  3 &
  84.9$\pm$0.9 &
  79.0$\pm$1.1 &
  79.5$\pm$0.7 &
  73.9$\pm$0.5 \\ \hline
\multirow{7}{*}{2023} &
  Wilkinghoff~\cite{Wilkinghoff2024b} &
  1 &
  71.2$\pm$1.6 &
  75.0$\pm$1.5 &
  75.5$\pm$0.8 &
  68.7$\pm$2.2 \\
 &
  \texttt{Ba}~\cite{FujimuraNU2024} &
  1 &
  72.0$\pm$1.4 &
  74.7$\pm$1.5 &
  78.0$\pm$1.5 &
  68.3$\pm$2.1 \\
 &
  \texttt{Ba}+\(\mathcal{L}_{\text{trp}}\) &
  1 &
  70.5$\pm$2.8 &
  74.3$\pm$2.4 &
  71.5$\pm$1.7 &
  66.4$\pm$1.2 \\
 &
  \texttt{Ba}+\texttt{Ex} &
  2 &
  \textbf{72.1$\pm$1.3} &
  \textbf{77.2$\pm$0.8} &
  \textbf{79.0$\pm$0.3} &
  \textbf{69.2$\pm$1.7} \\
 &
  \texttt{Ba}+\texttt{Ps} &
  2 &
  71.3$\pm$1.0 &
  \textbf{77.2$\pm$1.4} &
  75.5$\pm$2.4 &
  67.3$\pm$2.1 \\
 &
  \texttt{Ba}+\texttt{Ps}+\(\mathcal{L}_{\text{trp}}\) &
  2 &
  69.4$\pm$0.7 &
  74.9$\pm$1.2 &
  70.9$\pm$1.9 &
  67.3$\pm$2.5 \\
 &
  \texttt{Ba}+\texttt{Ex} &
  3 &
  70.9$\pm$0.8 &
  76.6$\pm$1.4 &
  \textbf{79.0$\pm$1.1} &
  69.0$\pm$1.6 \\ \hline
\multirow{7}{*}{2024} &
  Wilkinghoff~\cite{Wilkinghoff2024b} &
  1 &
  68.9$\pm$1.3 &
  63.8$\pm$1.8 &
  63.2$\pm$1.4 &
  62.7$\pm$1.7 \\
 &
  \texttt{Ba}~\cite{FujimuraNU2024} &
  1 &
  74.6$\pm$1.1 &
  64.5$\pm$1.8 &
  64.0$\pm$1.4 &
  65.3$\pm$2.4 \\
 &
  \texttt{Ba}+\(\mathcal{L}_{\text{trp}}\) &
  1 &
  70.9$\pm$1.4 &
  65.4$\pm$0.7 &
  60.9$\pm$2.9 &
  64.7$\pm$1.9 \\
 &
  \texttt{Ba}+\texttt{Ex} &
  2 &
  \textbf{75.2$\pm$0.2} &
  \textbf{65.5$\pm$1.5} &
  \textbf{64.6$\pm$2.2} &
  \textbf{67.3$\pm$1.0} \\
 &
  \texttt{Ba}+\texttt{Ps} &
  2 &
  72.5$\pm$0.7 &
  \textbf{65.5$\pm$1.9} &
  60.9$\pm$1.1 &
  63.0$\pm$2.0 \\
 &
  \texttt{Ba}+\texttt{Ps}+\(\mathcal{L}_{\text{trp}}\) &
  2 &
  71.3$\pm$1.9 &
  64.9$\pm$1.7 &
  59.6$\pm$1.5 &
  64.7$\pm$1.0 \\
 &
  \texttt{Ba}+\texttt{Ex} &
  3 &
  74.2$\pm$1.5 &
  64.7$\pm$1.2 &
  64.5$\pm$2.2 &
  67.1$\pm$0.8 \\ \hline
\end{tabular}
}
\end{table*}
\FloatBarrier                        
\section*{Biographies}
\noindent\textbf{Ibuki Kuroyanagi} received his M.E. degree in informatics from Nagoya University, Nagoya, Japan, in 2023. He is currently working toward an Ph.D. degree in informatics at Nagoya University. His research interests include audio signal processing. He is a student member of the Acoustical Society of Japan, and received the Acoustical Society of Japan 2021 Student Presentation Award.
In 2022, he received the DCASE Judges' Award.

\vspace{1em} 
\noindent\textbf{Takuya Fujimura} received his M.E. degree in informatics from Nagoya University, Nagoya, Japan, in 2024. He is currently working toward an Ph.D. degree in informatics at Nagoya University. His research interests include audio signal processing. He is a student member of the Acoustical Society of Japan.

\vspace{1em} 
\noindent\textbf{Kazuya Takeda} received the B.E. and M.E. degree in electrical engineering and the Dr.Eng. degree from Nagoya University, Nagoya, Japan, in 1983, 1985, and 1994, respectively.,From 1986 to 1989, he was with the Advanced Telecommunication Research (ATR) Laboratories, Osaka, Japan. His research interest at ATR was corpus-based speech synthesis. He was a Visiting Scientist with the Massachusetts Institute of Technology, Cambridge, from November 1987 to April 1988. from 1989 to 1995, he was a Researcher and Research Supervisor with KDD Research and Development Laboratories, Kamifukuoka, Japan. from 1995 to 2003, he was an Associate Professor with the Faculty of Engineering, Nagoya University. Since 2003, he has been a Professor with the Department of Media Science, Graduate School of Information Science, Nagoya University. He is an author or coauthor of more than 100 journal papers, six books, and more than 100 conference proceeding papers. His current research interests are media signal processing and its applications, including spatial audio, robust speech recognition, and driving behavior modeling.,Dr. Takeda was a Conference Technical Cochair of the International Conference on Multimodal Interfaces in 2007 and the International Conference on Vehicular Safety and Electronics in 2009. He was a co-founder of the Biennial Workshop on Digital Signal Processing for In-Vehicle Systems and Safety in 2003.

\vspace{1em} 
\noindent\textbf{Tomoki Toda} received the B.E. degree from Nagoya University, Nagoya, Japan, in 1999, and the M.E. and D.E. degrees from Nara Institute of Science and Technology (NAIST), Ikoma, Japan, in 2001 and 2003, respectively. He was a Research Fellow with the Japan Society for the Promotion of Science, from 2003 to 2005. He was then an Assistant Professor (2005–2011) and an Associate Professor (2011–2015) at NAIST. From 2015, he has been a Professor with the Information Technology Center, Nagoya University. His research interests include statistical approaches to speech processing. He was the recipient of more than 10 paper/achievement awards including the IEEE SPS 2009 Young Author Best Paper Award and the 2013 EURASIP-ISCA Best Paper Award (Speech Communication Journal).
\section*{References}
\printbibliography

@article{Chen2018,
    author={Chen, Baotong and Wan, Jiafu and Shu, Lei and Li, Peng and Mukherjee, Mithun and Yin, Boxing},
    journal={IEEE Access}, 
    title={{Smart Factory of Industry 4.0: Key Technologies, Application Case, and Challenges}}, 
    year={2018},
    volume={6},
    number={},
    pages={6505--6519},
    doi={10.1109/ACCESS.2017.2783682}
}

@inproceedings{hayashi2018anomalous,
    title={{Anomalous sound event detection based on WavNnet}},
    author={Hayashi, Tomoki and Komatsu, Tatsuya and Kondo, Reishi and Toda, Tomoki and Takeda, Kazuya},
    booktitle={2018 26th European Signal Processing Conference (EUSIPCO)},
    pages={2494--2498},
    year={2018},
    organization={IEEE}
}

@inproceedings{Kawaguchi2019icassp,
    author={Kawaguchi, Yohei and Tanabe, Ryo and Endo, Takashi and Ichige, Kenji and Hamada, Koichi},
    booktitle={ICASSP 2019 - 2019 IEEE International Conference on Acoustics, Speech and Signal Processing (ICASSP)}, 
    title={{Anomaly Detection Based on an Ensemble of Dereverberation and Anomalous Sound Extraction}}, 
    year={2019},
    volume={},
    number={},
    pages={865--869},
    doi={10.1109/ICASSP.2019.8683702}
}

@article{ad_survey2009,
    author = {Chandola, Varun and Banerjee, Arindam and Kumar, Vipin},
    title = {{Anomaly Detection: A Survey}},
    issue_date = {July 2009},
    publisher = {Association for Computing Machinery},
    address = {New York, NY, USA},
    volume = {41},
    number = {3},
    issn = {0360-0300},
    url = {https://doi.org/10.1145/1541880.1541882},
    doi = {10.1145/1541880.1541882},
    journal = {ACM Computing Surveys},
    month = {jul},
    articleno = {15},
    pages = {},
    note = {58 pages},
    year = {2009}
}

@inproceedings{Giri2020b,
    author = {Giri, Ritwik and Cheng, Fangzhou and Helwani, Karim and Tenneti, Srikanth V. and Isik, Umut and Krishnaswamy, Arvindh},
    title = {{Group Masked Autoencoder Based Density Estimator for Audio Anomaly Detection}},
    booktitle = {Proceedings of the Detection and Classification of Acoustic Scenes and Events 2020 Workshop (DCASE2020)},
    address = {Tokyo, Japan},
    month = {November},
    year = {2020},
    pages = {51--55},
}

@inproceedings{Purohit2020,
    author = {Purohit, Harsh and Tanabe, Ryo and Endo, Takashi and Suefusa, Kaori and Nikaido, Yuki and Kawaguchi, Yohei},
    title = {{Deep Autoencoding GMM-Based Unsupervised Anomaly Detection in Acoustic Signals and its Hyper-Parameter Optimization}},
    booktitle = {Proceedings of the Detection and Classification of Acoustic Scenes and Events 2020 Workshop (DCASE2020)},
    address = {Tokyo, Japan},
    month = {November},
    year = {2020},
    pages ={175--179},
}

@inproceedings{Kapka2020,
    author = {Kapka, Sławomir},
    title = {{ID-Conditioned Auto-Encoder for Unsupervised Anomaly Detection}},
    booktitle = {Proceedings of the Detection and Classification of Acoustic Scenes and Events 2020 Workshop (DCASE2020)},
    address = {Tokyo, Japan},
    month = {November},
    year = {2020},
    pages = {71--75},
}

@article{XIA2022497,
    author = {Xia, Xuan and Pan, Xizhou and Li, Nan and He, Xing and Ma, Lin and Zhang, Xiaoguang and Ding, Ning},
    title = {{GAN-Based Anomaly Detection: A Review}},
    year = {2022},
    issue_date = {Jul 2022},
    publisher = {Elsevier Science Publishers B. V.},
    address = {NLD},
    volume = {493},
    number = {C},
    issn = {0925-2312},
    url = {https://doi.org/10.1016/j.neucom.2021.12.093},
    doi = {10.1016/j.neucom.2021.12.093},
    journal = {Neurocomputing},
    month = {jul},
    pages = {497-–535},
    numpages = {39}
}

@inproceedings{Mishra2021,
    author={Mishra, Pankaj and Verk, Riccardo and Fornasier, Daniele and Piciarelli, Claudio and Foresti, Gian Luca},
    booktitle={2021 IEEE 30th International Symposium on Industrial Electronics (ISIE)}, 
    title={{VT-ADL: A Vision Transformer Network for Image Anomaly Detection and Localization}}, 
    year={2021},
    volume={},
    number={},
    pages={1--6},
    doi={10.1109/ISIE45552.2021.9576231}
}

@inproceedings{NEURIPS2020_NF,
    author = {Kirichenko, Polina and Izmailov, Pavel and Wilson, Andrew G},
    booktitle = {Proc. International Conference on Neural Information Processing Systems},
    pages = {20578--20589},
    publisher = {Curran Associates, Inc.},
    title = {{Why Normalizing Flows Fail to Detect Out-of-Distribution Data}},
    url = {https://proceedings.neurips.cc/paper/2020/file/ecb9fe2fbb99c31f567e9823e884dbec-Paper.pdf},
    year = {2020}
}

@inproceedings{Dohi2021icassp,
    author={Dohi, Kota and Endo, Takashi and Purohit, Harsh and Tanabe, Ryo and Kawaguchi, Yohei},
    booktitle={ICASSP 2021 - 2021 IEEE International Conference on Acoustics, Speech and Signal Processing (ICASSP)}, 
    title={{Flow-Based Self-Supervised Density Estimation for Anomalous Sound Detection}}, 
    year={2021},
    volume={},
    number={},
    pages={336--340},
    doi={10.1109/ICASSP39728.2021.9414662}
}

@INPROCEEDINGS{GuanJian2023,
  author={Guan, Jian and Liu, Youde and Zhu, Qiaoxi and Zheng, Tieran and Han, Jiqing and Wang, Wenwu},
  booktitle={ICASSP 2023 - 2023 IEEE International Conference on Acoustics, Speech and Signal Processing (ICASSP)}, 
  title={{Time-Weighted Frequency Domain Audio Representation with GMM Estimator for Anomalous Sound Detection}}, 
  year={2023},
  volume={},
  number={},
  pages={1-5},
  doi={10.1109/ICASSP49357.2023.10096356}}

@INPROCEEDINGS{JiangAnbai2023,
  author={Jiang, Anbai and Zhang, Wei-Qiang and Deng, Yufeng and Fan, Pingyi and Liu, Jia},
  booktitle={ICASSP 2023 - 2023 IEEE International Conference on Acoustics, Speech and Signal Processing (ICASSP)}, 
  title={{Unsupervised Anomaly Detection and Localization of Machine Audio: A Gan-Based Approach}}, 
  year={2023},
  volume={},
  number={},
  pages={1-5},
  doi={10.1109/ICASSP49357.2023.10096813}}

@inproceedings{Wilkinghoff2021b,
    author = {Wilkinghoff, Kevin},
    title = {{Combining Multiple Distributions based on Sub-Cluster AdaCos for Anomalous Sound Detection under Domain Shifted Conditions}},
    booktitle = {Proceedings of the 6th Detection and Classification of Acoustic Scenes and Events 2021 Workshop (DCASE2021)},
    year = {2021},
    pages = {55--59},
    isbn = {978-84-09-36072-7},
    doi. = {10.5281/zenodo.5770113}
}

@INPROCEEDINGS{Wilkinghoff2023,
  author={Wilkinghoff, Kevin},
  booktitle={ICASSP 2023 - 2023 IEEE International Conference on Acoustics, Speech and Signal Processing (ICASSP)}, 
  title={{Design Choices for Learning Embeddings from Auxiliary Tasks for Domain Generalization in Anomalous Sound Detection}}, 
  year={2023},
  volume={},
  number={},
  pages={1-5},
  doi={10.1109/ICASSP49357.2023.10097176}}

@INPROCEEDINGS{Hojjati2022,
  author={Hojjati, Hadi and Armanfard, Narges},
  booktitle={ICASSP 2022 - 2022 IEEE International Conference on Acoustics, Speech and Signal Processing (ICASSP)}, 
  title={{Self-Supervised Acoustic Anomaly Detection Via Contrastive Learning}}, 
  year={2022},
  volume={},
  number={},
  pages={3253-3257},
  doi={10.1109/ICASSP43922.2022.9746207}}

@INPROCEEDINGS{Guan2023,
  author={Guan, Jian and Xiao, Feiyang and Liu, Youde and Zhu, Qiaoxi and Wang, Wenwu},
  booktitle={ICASSP 2023 - 2023 IEEE International Conference on Acoustics, Speech and Signal Processing (ICASSP)}, 
  title={{Anomalous Sound Detection Using Audio Representation with Machine ID Based Contrastive Learning Pretraining}}, 
  year={2023},
  volume={},
  number={},
  pages={1-5},
  doi={10.1109/ICASSP49357.2023.10096054}}

@INPROCEEDINGS{Chen2023wavenet,
  author={Chen, Haihui and Ran, Likai and Sun, Xixia and Cai, Chao},
  booktitle={ICASSP 2023 - 2023 IEEE International Conference on Acoustics, Speech and Signal Processing (ICASSP)}, 
  title={{SW-WAVENET: Learning Representation from Spectrogram and Wavegram Using Wavenet for Anomalous Sound Detection}}, 
  year={2023},
  volume={},
  number={},
  pages={1-5},
  doi={10.1109/ICASSP49357.2023.10096742}}

@INPROCEEDINGS {Yao2023,
author = {X. Yao and R. Li and J. Zhang and J. Sun and C. Zhang},
booktitle = {2023 IEEE/CVF Conference on Computer Vision and Pattern Recognition (CVPR)},
title = {{Explicit Boundary Guided Semi-Push-Pull Contrastive Learning for Supervised Anomaly Detection}},
year = {2023},
volume = {},
issn = {},
pages = {24490-24499},
doi = {10.1109/CVPR52729.2023.02346},
url = {https://doi.ieeecomputersociety.org/10.1109/CVPR52729.2023.02346},
publisher = {IEEE Computer Society},
address = {Los Alamitos, CA, USA},
month = {jun}
}

@inproceedings{Koizumi_DCASE2020_01,
    Author = {Koizumi, Yuma and Kawaguchi, Yohei and Imoto, Keisuke and Nakamura, Toshiki and Nikaido, Yuki and Tanabe, Ryo and Purohit, Harsh and Suefusa, Kaori and Endo, Takashi and Yasuda, Masahiro and Harada, Noboru},
    title = {{Description and Discussion on {DCASE}2020 Challenge Task2: Unsupervised Anomalous Sound Detection for Machine Condition Monitoring}},
    year = {2020},
    booktitle = {Proceedings of the Detection and Classification of Acoustic Scenes and Events 2020 Workshop (DCASE2020)},
    month = {November},
    pages = {81--85},
}

@inproceedings{Kawaguchi2021,
    author = {Kawaguchi, Yohei and Imoto, Keisuke and Koizumi, Yuma and Harada, Noboru and Niizumi, Daisuke and Dohi, Kota and Tanabe, Ryo and Purohit, Harsh and Endo, Takashi},
    title = {{Description and Discussion on DCASE 2021 Challenge Task 2: Unsupervised Anomalous Detection for Machine Condition Monitoring Under Domain Shifted Conditions}},
    booktitle = {Proceedings of the 6th Detection and Classification of Acoustic Scenes and Events 2021 Workshop (DCASE2021)},
    address = {Barcelona, Spain},
    month = {November},
    year = {2021},
    pages = {186--190},
}

@inproceedings{Koizumi_WASPAA2019_01,
    Author = {Koizumi, Yuma and Saito, Shoichiro and Uematsu, Hisashi and Harada, Noboru and Imoto, Keisuke},
    title = {{ToyADMOS: A Dataset of Miniature-machine Operating Sounds for Anomalous Sound Detection}},
    year = {2019},
    booktitle = {Proceedings of {IEEE} Workshop on Applications of Signal Processing to Audio and Acoustics ({WASPAA})},
    month = {November},
    pages = {308--312},
    url = {https://ieeexplore.ieee.org/document/8937164}
}

@inproceedings{Purohit_DCASE2019_01,
    Author = {Purohit, Harsh and Tanabe, Ryo and Ichige, Takeshi and Endo, Takashi and Nikaido, Yuki and Suefusa, Kaori and Kawaguchi, Yohei},
    title = {{MIMII Dataset: Sound Dataset for Malfunctioning Industrial Machine Investigation and Inspection}},
    year = {2019},
    booktitle = {Proceedings of the Detection and Classification of Acoustic Scenes and Events 2019 Workshop ({DCASE2019})},
    month = {November},
    pages = {209--213},
    url = {http://dcase.community/documents/workshop2019/proceedings/DCASE2019Workshop\_Purohit\_21.pdf}
}

@Inbook{Reynolds2009,
    author={Reynolds, Douglas},
    title={{Gaussian Mixture Models}},
    bookTitle={Encyclopedia of Biometrics},
    year={2009},
    publisher={Springer US},
    address={Boston, MA},
    pages={659--663},
    isbn={978-0-387-73003-5},
    doi={10.1007/978-0-387-73003-5_196},
    url={https://doi.org/10.1007/978-0-387-73003-5_196}
}

@inproceedings{liu2019,
    author={Liu, Wenbo and Cui, Delong and Peng, Zhiping and Zhong, Jihai},
    booktitle={2019 IEEE International Conference on Power, Intelligent Computing and Systems (ICPICS)}, 
    title={{Outlier Detection Algorithm Based on Gaussian Mixture Model}}, 
    year={2019},
    volume={},
    number={},
    pages={488--492},
    doi={10.1109/ICPICS47731.2019.8942474}
}

@inproceedings{uefusa2020,
    author={Suefusa, Kaori and Nishida, Tomoya and Purohit, Harsh and Tanabe, Ryo and Endo, Takashi and Kawaguchi, Yohei},
    booktitle={ICASSP 2020 - 2020 IEEE International Conference on Acoustics, Speech and Signal Processing (ICASSP)}, 
    title={{Anomalous Sound Detection Based on Interpolation Deep Neural Network}}, 
    year={2020},
    volume={},
    number={},
    pages={271--275},
    doi={10.1109/ICASSP40776.2020.9054344}
}

@inproceedings{Ruff2020Deep,
    title={{Deep Semi-Supervised Anomaly Detection}},
    author={Lukas Ruff and Robert A. Vandermeulen and Nico Görnitz and Alexander Binder and Emmanuel M{\"u}ller and Klaus-Robert M{\"u}ller and Marius Kloft},
    booktitle={International Conference on Learning Representations},
    year={2020},
    url={https://openreview.net/forum?id=HkgH0TEYwH},
    note={23 pages},
}

@inproceedings{ruff2020rethinking,
     title     = {{Rethinking Assumptions in Deep Anomaly Detection}},
     author    = {Ruff, Lukas and Vandermeulen, Robert A and Franks, Billy Joe and M{\"u}ller, Klaus-Robert and Kloft, Marius},
     booktitle = {ICML 2021 Workshop on Uncertainty \& Robustness in Deep Learning},
     year      = {2021}
}

@inproceedings{kuroyanagi2021anomalous,
    title={{Anomalous Sound Detection Using a Binary Classification Model and Class Centroids}}, 
    author={Ibuki Kuroyanagi and Tomoki Hayashi and Kazuya Takeda and Tomoki Toda},
    booktitle={2021 29th European Signal Processing Conference (EUSIPCO)},
    pages={1995--1999},
    year={2021},
    organization={IEEE}
}

@inproceedings{stgrammfn2022,
    author={Liu, Youde and Guan, Jian and Zhu, Qiaoxi and Wang, Wenwu},
    booktitle={ICASSP 2022 - 2022 IEEE International Conference on Acoustics, Speech and Signal Processing (ICASSP)}, 
    title={{Anomalous Sound Detection Using Spectral-Temporal Information Fusion}}, 
    year={2022},
    volume={},
    number={},
    pages={816--820},
    doi={10.1109/ICASSP43922.2022.9747868}
}

@inproceedings{primus2020anomalous,
    author = {Primus, Paul and Haunschmid, Verena and Praher, Patrick and Widmer, Gerhard},
    title = {{Anomalous Sound Detection as a Simple Binary Classification Problem with Careful Selection of Proxy Outlier Examples}},
    booktitle = {Proceedings of the Detection and Classification of Acoustic Scenes and Events 2020 Workshop (DCASE2020)},
    address = {Tokyo, Japan},
    month = {November},
    year = {2020},
    pages = {170--174},
}

@INPROCEEDINGS {ding2022catching,
author = {C. Ding and G. Pang and C. Shen},
booktitle = {2022 IEEE/CVF Conference on Computer Vision and Pattern Recognition (CVPR)},
title = {{Catching Both Gray and Black Swans: Open-set Supervised Anomaly Detection}},
year = {2022},
volume = {},
issn = {},
pages = {7378-7388},
doi = {10.1109/CVPR52688.2022.00724},
url = {https://doi.ieeecomputersociety.org/10.1109/CVPR52688.2022.00724},
publisher = {IEEE Computer Society},
address = {Los Alamitos, CA, USA},
month = {jun}
}

@inproceedings{Kuroyanagi2021dcasew,
    author = {Kuroyanagi, Ibuki and Hayashi, Tomoki and Adachi, Yusuke and Yoshimura, Takenori and Takeda, Kazuya and Toda, Tomoki},
    title = {{An Ensemble Approach to Anomalous Sound Detection Based on Conformer-Based Autoencoder and Binary Classifier Incorporated with Metric Learning}},
    booktitle = {Proceedings of the 6th Detection and Classification of Acoustic Scenes and Events 2021 Workshop (DCASE2021)},
    address = {Barcelona, Spain},
    month = {November},
    year = {2021},
    pages = {110--114},
}

@techreport{KuroyanagiNUHDL2022,
    Author = {Kuroyanagi, Ibuki and Hayashi, Tomoki and Takeda, Kazuya and Toda, Tomoki},
    title = {{Two-stage anomalous sound detection systems using domain generalization and specialization techniques}},
    institution = {DCASE2022 Challenge},
    year = {2022},
    month = {July},
    note={5 pages},
}

@inproceedings{NIPS2017_6c1da886,
    author = {Papamakarios, George and Pavlakou, Theo and Murray, Iain},
    title = {{Masked Autoregressive Flow for Density Estimation}},
    year = {2017},
    isbn = {9781510860964},
    publisher = {Curran Associates Inc.},
    address = {Red Hook, NY, USA},
    booktitle = {Proc. International Conference on Neural Information Processing Systems},
    pages = {2335–-2344},
    numpages = {10},
    series = {NIPS'17}
}

@inproceedings{Dohi2022_2,
    author = {Dohi, Kota and Imoto, Keisuke and Harada, Noboru and Niizumi, Daisuke and Koizumi, Yuma and Nishida, Tomoya and Purohit, Harsh and Tanabe, Ryo and Endo, Takashi and Yamamoto, Masaaki and Kawaguchi, Yohei},
    title = {{Description and Discussion on DCASE 2022 Challenge Task 2: Unsupervised Anomalous Sound Detection for Machine Condition Monitoring Applying Domain Generalization Techniques}},
    booktitle = {Proceedings of the 7th Detection and Classification of Acoustic Scenes and Events 2022 Workshop (DCASE2022)},
    address = {Nancy, France},
    month = {November},
    year = {2022},
    note={5 pages},
}

@inproceedings{Dohi2022,
    author = {Dohi, Kota and Nishida, Tomoya and Purohit, Harsh and Tanabe, Ryo and Endo, Takashi and Yamamoto, Masaaki and Nikaido, Yuki and Kawaguchi, Yohei},
    title = {{MIMII DG: Sound Dataset for Malfunctioning Industrial Machine Investigation and Inspection for Domain Ggeneralization Task}},
    booktitle = {Proceedings of the 7th Detection and Classification of Acoustic Scenes and Events 2022 Workshop (DCASE2022)},
    address = {Nancy, France},
    month = {November},
    year = {2022},
    note={5 pages},
}

@inproceedings{Dohi2023,
    author = {Dohi, Kota and Imoto, Keisuke and Harada, Noboru and Niizumi, Daisuke and Koizumi, Yuma and Nishida, Tomoya and Purohit, Harsh and Tanabe, Ryo and Endo, Takashi and Kawaguchi, Yohei},
    title = {{Description and Discussion on DCASE 2023 Challenge Task 2: First-Shot Unsupervised Anomalous Sound Detection for Machine Condition Monitoring}},
    booktitle = {Proceedings of the 8th Detection and Classification of Acoustic Scenes and Events 2023 Workshop (DCASE2023)},
    address = {Tampere, Finland},
    month = {September},
    year = {2023},
    pages = {31--35},
}

@inproceedings{mixupzhang2018,
    title={{Mixup: Beyond Empirical Risk Minimization}},
    author={Hongyi Zhang and Moustapha Cisse and Yann N. Dauphin and David Lopez-Paz},
    booktitle={International Conference on Learning Representations},
    year={2018},
    url={https://openreview.net/forum?id=r1Ddp1-Rb},
    note={13 pages},
}

@inproceedings{adamw2019,
    author    = {Ilya Loshchilov and Frank Hutter},
    title     = {{Decoupled Weight Decay Regularization}},
    booktitle = {7th International Conference on Learning Representations},
    publisher = {OpenReview.net},
    year      = {2019},
    url       = {https://openreview.net/forum?id=Bkg6RiCqY7},
    bibsource = {dblp computer science bibliography, https://dblp.org},
    note={8 pages},
}

@INPROCEEDINGS{geco2023,
  author={Zeng, Xiao-Min and Song, Yan and Zhuo, Zhu and Zhou, Yu and Li, Yu-Hong and Xue, Hui and Dai, Li-Rong and McLoughlin, Ian},
  booktitle={ICASSP 2023 - 2023 IEEE International Conference on Acoustics, Speech and Signal Processing (ICASSP)}, 
  title={Joint Generative-Contrastive Representation Learning for Anomalous Sound Detection}, 
  year={2023},
  volume={},
  number={},
  pages={1--5},
  doi={10.1109/ICASSP49357.2023.10095568}}

@ARTICLE{Wilkinghoff2024,
  author={Wilkinghoff, Kevin and Kurth, Frank},
  journal={IEEE/ACM Transactions on Audio, Speech, and Language Processing}, 
  title={{Why Do Angular Margin Losses Work Well for Semi-Supervised Anomalous Sound Detection?}}, 
  year={2024},
  volume={32},
  number={},
  pages={608--622},
  doi={10.1109/TASLP.2023.3337153},
  ISSN={2329-9304},
  month={},}

@INPROCEEDINGS{Wilkinghoff2024b,
  author={Wilkinghoff, Kevin},
  booktitle={ICASSP 2024 - 2024 IEEE International Conference on Acoustics, Speech and Signal Processing (ICASSP)}, 
  title={{Self-Supervised Learning for Anomalous Sound Detection}}, 
  year={2024},
  volume={},
  number={},
  pages={276-280},
  doi={10.1109/ICASSP48485.2024.10447156}}

@inproceedings{Nishida2024,
    author = {Nishida, Tomoya and Harada, Noboru and Niizumi, Daisuke and Albertini, Davide and Sannino, Roberto and Pradolini, Simone and Augusti, Filippo and Imoto, Keisuke and Dohi, Kota and Purohit, Harsh and Endo, Takashi and Kawaguchi, Yohei},
    title = {{Description and Discussion on DCASE 2024 Challenge Task 2: First-Shot Unsupervised Anomalous Sound Detection for Machine Condition Monitoring}},
    booktitle = {Proceedings of the Detection and Classification of Acoustic Scenes and Events 2024 Workshop (DCASE2024)},
    month = {October},
    year = {2024},
    pages = {111--115},
}

@inproceedings{choi2024noisy,
  title={{Noisy-Arcmix: Additive Noisy Angular Margin Loss Combined With Mixup For Anomalous Sound Detection}},
  author={Choi, Soonhyeon and Choi, Jung-Woo},
  booktitle={ICASSP 2024-2024 IEEE International Conference on Acoustics, Speech and Signal Processing (ICASSP)},
  pages={516--520},
  year={2024},
  organization={IEEE}
}

@inproceedings{Harada2021,
    author = {Harada, Noboru and Niizumi, Daisuke and Takeuchi, Daiki and Ohishi, Yasunori and Yasuda, Masahiro and Saito, Shoichiro},
    title = {{ToyADMOS2: Another Dataset of Miniature-Machine Operating Sounds for Anomalous Sound Detection under Domain Shift Conditions}},
    booktitle = {Proceedings of the 6th Detection and Classification of Acoustic Scenes and Events 2021 Workshop (DCASE2021)},
    address = {Barcelona, Spain},
    month = {November},
    year = {2021},
    pages = {1--5},
    isbn = {978-84-09-36072-7},
    doi. = {10.5281/zenodo.5770113}
}

@inproceedings{Dohi2022_mimii,
    author = {Dohi, Kota and Nishida, Tomoya and Purohit, Harsh and Tanabe, Ryo and Endo, Takashi and Yamamoto, Masaaki and Nikaido, Yuki and Kawaguchi, Yohei},
    title = {{MIMII DG: Sound Dataset for Malfunctioning Industrial Machine Investigation and Inspection for Domain Generalization Task}},
    booktitle = {Proceedings of the 7th Detection and Classification of Acoustic Scenes and Events 2022 Workshop (DCASE2022)},
    address = {Nancy, France},
    month = {November},
    year = {2022},
    pages = {1--5},
}

@article{Tanabe_WASPAA2021_01,
    author = {Tanabe, Ryo and Purohit, Harsh and Dohi, Kota and Endo, Takashi and Nikaido, Yuki and Nakamura, Toshiki and Kawaguchi, Yohei},
    title = {{MIMII DUE: Sound Dataset for Malfunctioning Industrial Machine Investigation and Inspection with Domain Shifts due to Changes in Operational and Environmental Conditions}},
    journal = {IEEE Workshop on Applications of Signal Processing to Audio and Acoustics (WASPAA)},
    year = {2021},
    pages = {21--25},
    doi = {10.1109/WASPAA52581.2021.9632802}
}

@article{Harada_EUSIPCO2023_01,
    author = {Harada, Noboru and Niizumi, Daisuke and Takeuchi, Daiki and Ohishi, Yasunori and Yasuda, Masahiro},
    title = {{First-Shot Anomaly Detection for Machine Condition Monitoring: A Domain Generalization Baseline}},
    journal = {Proceedings of 31st European Signal Processing Conference (EUSIPCO)},
    pages = {191--195},
    year = {2023}
}

@INPROCEEDINGS{audioset,
  author={Gemmeke, Jort F. and Ellis, Daniel P. W. and Freedman, Dylan and Jansen, Aren and Lawrence, Wade and Moore, R. Channing and Plakal, Manoj and Ritter, Marvin},
  booktitle={2017 IEEE International Conference on Acoustics, Speech and Signal Processing (ICASSP)}, 
  title={{Audio Set: An ontology and human-labeled dataset for audio events}}, 
  year={2017},
  volume={},
  number={},
  pages={776-780},
  doi={10.1109/ICASSP.2017.7952261}}

@INPROCEEDINGS{Wilkinghoff2023eu,
  author={Wilkinghoff, Kevin and Fritz, Fabian},
  booktitle={2023 31st European Signal Processing Conference (EUSIPCO)}, 
  title={{On Using Pre-Trained Embeddings for Detecting Anomalous Sounds with Limited Training Data}}, 
  year={2023},
  volume={},
  number={},
  pages={186-190},
  doi={10.23919/EUSIPCO58844.2023.10290003}}

@techreport{FujimuraNU2024,
    Author = {Takuya, Fujimura and Kuroyanagi, Ibuki and Toda, Tomoki},
    title = {{The NU systems for DCASE 2024 Challenge Task 2}},
    institution = {DCASE2024 Challenge},
    year = {2024},
    note={5 pages}
}

@inproceedings{Harada2023,
    author = {Harada, Noboru and Niizumi, Daisuke and Takeuchi, Daiki and Ohishi, Yasunori and Yasuda, Masahiro},
    title = {{ToyADMOS2+: New Toyadmos Data and Benchmark Results of the First-Shot Anomalous Sound Event Detection Baseline}},
    booktitle = {Proceedings of the 8th Detection and Classification of Acoustic Scenes and Events 2023 Workshop (DCASE2023)},
    address = {Tampere, Finland},
    month = {September},
    year = {2023},
    pages = {41--45},
}

@inproceedings{Albertini2024,
    author = {Albertini, Davide and Augusti, Filippo and Esmer, Kudret and Bernardini, Alberto and Sannino, Roberto},
    title = {{IMAD-DS: A Dataset for Industrial Multi-Sensor Anomaly Detection Under Domain Shift Conditions}},
    booktitle = {Proceedings of the Detection and Classification of Acoustic Scenes and Events 2024 Workshop (DCASE2024)},
    address = {Tokyo, Japan},
    month = {October},
    year = {2024},
    pages = {1--5},
}

@inproceedings{Niizumi2024,
    author = {Niizumi, Daisuke and Harada, Noboru and Ohishi, Yasunori and Takeuchi, Daiki and Yasuda, Masahiro},
    title = {{ToyADMOS2\#: Yet Another Dataset for the DCASE2024 Challenge Task 2 First-Shot Anomalous Sound Detection}},
    booktitle = {Proceedings of the Detection and Classification of Acoustic Scenes and Events 2024 Workshop (DCASE2024)},
    address = {Tokyo, Japan},
    month = {October},
    year = {2024},
    pages = {106--110},
}

@article{auc_Donna,
author = {Donna Katzman McClish},
title ={{Analyzing a Portion of the ROC Curve}},
journal = {Medical Decision Making},
volume = {9},
number = {3},
pages = {190--195},
year = {1989},
doi = {10.1177/0272989X8900900307},
URL = { 
        https://doi.org/10.1177/0272989X8900900307
},
}

@INPROCEEDINGS{auc_Ebbers,
  author={Ebbers, Janek and Haeb-Umbach, Reinhold and Serizel, Romain},
  booktitle={ICASSP 2022 - 2022 IEEE International Conference on Acoustics, Speech and Signal Processing (ICASSP)}, 
  title={{Threshold Independent Evaluation of Sound Event Detection Scores}}, 
  year={2022},
  volume={},
  number={},
  pages={1021-1025},
  doi={10.1109/ICASSP43922.2022.9747556}}

@ARTICLE{Wang2023,
  author={Wang, Jindong and Lan, Cuiling and Liu, Chang and Ouyang, Yidong and Qin, Tao and Lu, Wang and Chen, Yiqiang and Zeng, Wenjun and Yu, Philip S.},
  journal={IEEE Transactions on Knowledge and Data Engineering}, 
  title={{Generalizing to Unseen Domains: A Survey on Domain Generalization}}, 
  year={2023},
  volume={35},
  number={8},
  pages={8052-8072},
  doi={10.1109/TKDE.2022.3178128}}

@INPROCEEDINGS{torchaudio2023,
  author={Hwang, Jeff and Hira, Moto and Chen, Caroline and Zhang, Xiaohui and Ni, Zhaoheng and Sun, Guangzhi and Ma, Pingchuan and Huang, Ruizhe and Pratap, Vineel and Zhang, Yuekai and Kumar, Anurag and Yu, Chin-Yun and Zhu, Chuang and Liu, Chunxi and Kahn, Jacob and Ravanelli, Mirco and Sun, Peng and Watanabe, Shinji and Shi, Yangyang and Tao, Yumeng},
  booktitle={2023 IEEE Automatic Speech Recognition and Understanding Workshop (ASRU)}, 
  title={{TorchAudio 2.1: Advancing Speech Recognition, Self-Supervised Learning, and Audio Processing Components for Pytorch}}, 
  year={2023},
  volume={},
  number={},
  pages={1-9},
  doi={10.1109/ASRU57964.2023.10389648}}

@INPROCEEDINGS{torchaudio2022,
  author={Yang, Yao-Yuan and Hira, Moto and Ni, Zhaoheng and Astafurov, Artyom and Chen, Caroline and Puhrsch, Christian and Pollack, David and Genzel, Dmitriy and Greenberg, Donny and Yang, Edward Z. and Lian, Jason and Hwang, Jeff and Chen, Ji and Goldsborough, Peter and Narenthiran, Sean and Watanabe, Shinji and Chintala, Soumith and Quenneville-Bélair, Vincent},
  booktitle={ICASSP 2022 - 2022 IEEE International Conference on Acoustics, Speech and Signal Processing (ICASSP)}, 
  title={{Torchaudio: Building Blocks for Audio and Speech Processing}}, 
  year={2022},
  volume={},
  number={},
  pages={6982-6986},
  doi={10.1109/ICASSP43922.2022.9747236}}

@INPROCEEDINGS{Kuroyanagiicassp2022,
  author={Kuroyanagi, Ibuki and Komatsu, Tatsuya},
  booktitle={ICASSP 2022 - 2022 IEEE International Conference on Acoustics, Speech and Signal Processing (ICASSP)}, 
  title={Self-Supervised Learning Method Using Multiple Sampling Strategies for General-Purpose Audio Representation}, 
  year={2022},
  volume={},
  number={},
  pages={3263-3267},
  doi={10.1109/ICASSP43922.2022.9746798},
  ISSN={2379-190X},
  month={May},}

@INPROCEEDINGS{Fujimura2024eu,
  author={Fujimura, Takuya and Imoto, Keisuke and Toda, Tomoki},
  booktitle={2024 32nd European Signal Processing Conference (EUSIPCO)}, 
  title={{Discriminative Neighborhood Smoothing for Generative Anomalous Sound Detection}}, 
  year={2024},
  volume={},
  number={},
  pages={156-160},
  doi={10.23919/EUSIPCO63174.2024.10715201}}

@INPROCEEDINGS{Nishida2022eu,
  author={Nishida, Tomoya and Dohi, Kota and Endo, Takashi and Yamamoto, Masaaki and Kawaguchi, Yohei},
  booktitle={2022 30th European Signal Processing Conference (EUSIPCO)}, 
  title={{Anomalous Sound Detection Based on Machine Activity Detection}}, 
  year={2022},
  volume={},
  number={},
  pages={269-273},
  doi={10.23919/EUSIPCO55093.2022.9909901}}

@INPROCEEDINGS{fujimura2024ic,
  author={Fujimura, Takuya and Kuroyanagi, Ibuki and Toda, Tomoki},
  booktitle={ICASSP 2025 - 2025 IEEE International Conference on Acoustics, Speech and Signal Processing (ICASSP)}, 
  title={{Improvements of Discriminative Feature Space Training for Anomalous Sound Detection in Unlabeled Conditions}}, 
  year={2025},
  volume={},
  number={},
  pages={1-5},
  doi={10.1109/ICASSP49660.2025.10890020}}

@techreport{JiangCUP2024,
    Author = {Wang, Yaocong and Deng, Xinlong and Jiang, Jie and Kong, Qiuqiang},
    title = {{ANOMALOUS SOUND DETECTION BASED ON PSEUDO LABELS FROM GUIDED CLUSTERING}},
    institution = {DCASE2024 Challenge},
    year = {2024},
    note={3 pages}
}

@techreport{LvAITHU2024,
    Author = {Lv, Zhiqiang and Jiang, Anbai and Han, Bing and Liang, Yuzhe and Qian, Yanmin and Chen, Xie and Liu, Jia and Fan, Pingyi},
    title = {{AITHU System for First-Shot Unsupervised Anomalous Sound Detection}},
    institution = {DCASE2024 Challenge},
    year = {2024},
    note={4 pages}
}

@INPROCEEDINGS{Chen2022,
  author={Chen, Bingqing and Bondi, Luca and Das, Samarjit},
  booktitle={2022 26th International Conference on Pattern Recognition (ICPR)}, 
  title={Learning to Adapt to Domain Shifts with Few-shot Samples in Anomalous Sound Detection}, 
  year={2022},
  volume={},
  number={},
  pages={133-139},
  doi={10.1109/ICPR56361.2022.9956351}}

\end{document}